\documentclass[12pt,a4paper]{article}
\usepackage{epsfig}
\usepackage{amsfonts}
\usepackage{amssymb}
\usepackage{color}

\def\ep              {\varepsilon }
\def\Li              {\mbox{Li}}
\def\la              {\langle}
\def\ra              {\rangle}

\begin{document}
%
%
\begin{center}

\vspace{1cm}

{\bf \large ON SOFT THEOREMS AND FORM FACTORS IN $\mathcal{N}=4$ SYM THEORY.} \vspace{2cm}

{\bf \large L. V. Bork$^{1,2}$ A.I. Onishchenko$^{3,4}$}\vspace{0.5cm}

{\it $^1$Institute for Theoretical and Experimental Physics, Moscow,
Russia,\\
$^2$The Center for Fundamental and Applied Research, All-Russia
Research Institute of Automatics, Moscow, Russia, \\
$^3$Moscow Institute of Physics and Technology (State University), Dolgoprudny, Russia, \\
$^4$Skobeltsyn Institute of Nuclear Physics, Moscow State University, Moscow, Russia}\vspace{1cm}

\abstract{Soft theorems for the form factors of 1/2-BPS and Konishi
operator supermultiplets are derived at tree level in $\mathcal{N}=4$ SYM theory. They have a form identical to the one in the amplitude case. For MHV sectors of stress tensor and Konishi supermultiplets loop corrections to soft theorems are considered at one loop level. They also appear to have universal form in soft limit.  Possible generalization of the on-shell diagrams to the form factors based on leading soft behavior is suggested. Finally, we give some comments on inverse soft limit and integrability of form factors in the limit $q^2\to 0$ }
\end{center}


\newpage

\tableofcontents{}\vspace{0.5cm}

\renewcommand{\theequation}{\thesection.\arabic{equation}}
\section{Introduction}\label{p1}

In the last years there was a lot of progress in understanding the
structure of S-matrix (amplitudes) of four dimensional gauge theories. This was made mostly due to the development of new computational tools such as different sets of on shell recursion relations for tree level amplitudes and different unitarity based methods for loop amplitudes \cite{Reviews_Ampl_General,Henrietta_Amplitudes}.
The most impressive results were obtained in the theories with extended supersymmetry. It expected that full S-matrix of $\mathcal{N}=4$ super Yang-Mills (SYM) will be computed eventually in some form.

Using physical intuition and these new computational tools it became possible to find new structures even in such, seemingly well understood, areas as soft behavior of scattering amplitudes in four dimensional gauge theories and gravity. This way a new set of soft theorems was discovered. Soft theorems are in general a group of statements about universal factorization behavior of scattering amplitudes in gauge theories or gravity in the limit,  when one of the particle's momenta goes to zero.

Recent progress in the study of soft behavior of scattering amplitudes was made within the context of BMS \cite{BMS} and extended BMS \cite{extendedBMS} asymptotic symmetries  for asymptotically Minkowskian spacetimes. BMS and extended BMS transformations are infinitesimal diffeomorphisms, which preserve a prescribed structure of spacetime at future or past null infinities $\mathcal{I}^{\pm}$, but act nontrivially on the physical data residing there. $\mathcal{I}^{\pm}$ is given by the product of a conformal two-sphere (directions of outgoing or incoming null rays) with a null line (retarded times). The extended BMS algebra is the semi-direct sum of the infinite-dimensional Lie algebra of infinitesimal super-translations along null generators of $\mathcal{I}^+$ together with two copies of local Virasoro algebras given by conformal Killing vectors of the two-sphere, so called infinitesimal super-rotations. It turns out, that Weinberg's soft graviton theorem \cite{Weinberg_SoftGravitonTheorem} is actually equivalent to the Ward identity of BMS invariance (invariance under 'diagonal' subgroup of the product group of BMS super-translations acting on past and future null infinities) of the quantum gravity $\mathcal{S}$-matrix \cite{Stromingeк_BMS_GravitationalScattering,He_BMS_SoftTheorem}. Moreover, the invariance of quantum gravity $\mathcal{S}$-matrix under 'diagonal' subgroup of the past and future Virasoro symmetries, i.e. super-rotations, leads to another Ward identity connected with subleading soft behavior of gravitational scattering amplitudes \cite{Cachazo_NewSoftGravitonTheorem}. Further study in this direction clarified the connection between soft behavior of gravitational scattering amplitudes and gravitational memory effects, such as displacement memory and spin memory effects  \cite{Strominger_DisplacementMemory,Pasterski_SpinMemory}.

An analysis of asymptotic symmetries at future (past) null infinities $\mathcal{I}^{\pm}$ of Minkowski spacetime for Yang-Mills  theories with gauge group $\mathcal{G}$ resulted in the discovery of a infinite-dimensional $\mathcal{G}$ Kac-Moody symmetry of gauge theory $\mathcal{S}$-matrix \cite{Strominger_AsymptoticSymmetriesYM}. The corresponding  Ward identities are equivalent to gauge theory soft theorems \cite{Strominger_AsymptoticSymmetriesYM,He_KacMoodyYM}. Moreover, it was shown, that scattering amplitudes of any four-dimensional theory with nonabelian gauge group $\mathcal{G}$ may be recast as two-dimensional correlation functions on the asymptotic two-sphere at null infinity \cite{He_KacMoodyYM}.

The soft theorem for the  tree level color ordered amplitudes of $D=4$ YM theory can be written similar to the gravity case \cite{Cachazo_NewSoftGravitonTheorem}  as \cite{Laenen_PathIntegralEikonal,Casali_SoftSubLeadingYM}:
\begin{eqnarray}
A_{n+1}^{h_s,h_1,...,h_n}(\epsilon p_s,p_1,...,p_n)=
\left(\frac{\hat{S}_1}{\epsilon}+\hat{S}_2\right)A_{n}^{h_1,...,h_n}(p_1,...,p_n)
+O(\epsilon),~\epsilon \rightarrow0,
\end{eqnarray}
Here, $p_i$ - is the momentum of i'th particle (gluon in this case), $h_i$ - is its  helicity  and $p_s \mapsto \epsilon p_s$. The operators $\hat{S}_1$ and $\hat{S}_2$ are given by
\begin{eqnarray}
\hat{S}_1=\frac{(p_n\epsilon^{h_s}_s)}{\sqrt{2}(p_np_s)}+
\frac{(p_1\epsilon^{h_s}_s)}{\sqrt{2}(p_np_s)},
\end{eqnarray}
and
\begin{eqnarray}
\hat{S}_2=\frac{J^{\mu\nu}_n\epsilon^{h_s}_{s,\mu}p_{n,\nu}}{\sqrt{2}(p_np_s)}+
\frac{J^{\mu\nu}_1\epsilon^{h_s}_{s,\mu}p_{1,\nu}}{\sqrt{2}(p_1p_s)},
\end{eqnarray}
where $\epsilon^{h_i}_{i,\mu}$ is the polarization vector of $i$'th particle and $J^{\mu\nu}_i$ is the angular momentum generator acting on $i$'th particle. Within the spinor helicity formalism $\hat{S}_1$ and $\hat{S}_2$ operators can be compactly rewritten as
\begin{eqnarray}
\hat{S}_1=\frac{\langle 1n\rangle}{\langle ns\rangle\langle s1\rangle },
~\hat{S}_2=\frac{\tilde{\lambda}^{\dot{\alpha}}_s}{\langle s1 \rangle}\frac{\partial}{\partial\tilde{\lambda}^{\dot{\alpha}}_1}+
\frac{\tilde{\lambda}^{\dot{\alpha}}_s}{\langle sn \rangle}\frac{\partial}{\partial\tilde{\lambda}^{\dot{\alpha}}_n}.
\end{eqnarray}
As we already mentioned above, the form of $\hat{S}_1$ operator was known for quite a long time \cite{Weinberg_SoftGravitonTheorem}, while subleading universal behavior controlled by $\hat{S}_2$ operator was discovered only recently. At loop level soft theorems for amplitudes may, in general, receive radiative corrections \cite{Bern_Soft_YM_Grav_loops}
depending on the order of limits (loop regularization parameter $\varepsilon\rightarrow 0$ and then $p_s\rightarrow0$ or vice versa) \cite{Cachazo_Soft_YM_Grav_loops}.

There is another class of objects within gauge theory, which are very similar to the amplitudes - the form factors. It is known (at least in gauge theories with maximal supersymmetry), that form factors have properties, which are very similar to those of amplitudes and could be computed  with the use of similar methods as in the case of amplitudes.
It is natural to expect that soft theorems  will be  valid in some form for the form factors as well.

The purpose of this paper is to derive soft theorems for form factors of different sets of operators in $\mathcal{N}=4$ SYM at tree level and to study the structure of radiative corrections to the above tree-level soft theorems on some particular examples at one loop . We will also discuss the possible Grassmannian integral representation of form factors based on some insights from soft limit. It will be shown that in the limit of $q^2\to 0$ ($q$ is the form factor momentum) there is some evidence in favor of Yangian invariance of tree-level form factors corresponding to single trace operators studied previously in the context of $PSU(2,2|4)$ $\mathcal{N}=4$ SYM spin chain. Next, we will consider how the inverse soft limit (ISL) iterative procedure works in the case of form factors and how integrability and quantum inverse scattering method (QISM) could be used to construct Yangian invariants relevant for form factors in the limit of $q^2\to 0$.

The structure of this paper is the following. In section \ref{p2} we briefly
discuss the general structure of the form factors of the operators from
the 1/2-BPS and Konishi $\mathcal{N}=4$ SYM supermultiplets within on-shell harmonic and ordinary superspaces. In section \ref{p3} we derive soft theorems for general form factors of $p=2,3$
1/2-BPS operator supermultiplets at tree level. We also verify that soft theorems will likely hold for the form factors of general $p$ 1/2-BPS and Konishi operator supermultiplets on the known up to the moment
particular examples. Next we compute loop corrections to the soft theorems in the case of  MHV form factors of
$p=2$ 1/2-BPS and Konishi operator supermultiplets at one loop level. In section \ref{p4} we discuss possible generalization of  Grassmannian integral representation based on the on-shell diagrams for the case of form factors and present evidence in favor of Yangian invariance of tree-level form factors in the limit $q^2\to 0$. Section \ref{p5} contains the discussion of inverse soft limit iterative procedure both in the case of amplitudes and form factors together with QISM construction of Yangian invariants in these cases.

\section{Form factors in $\mathcal{N}=4$ SYM}\label{p2}

In general, a study of form factors in $\mathcal{N}=4$ SYM goes with  a consideration of form factors of operators from 1/2-BPS and Konishi supermultiplets. In this chapter we are going to introduce essential ideas and notation regarding these operator  supermultiplets formulated in different superspaces.

\subsection{Form factors of 1/2-BPS operator supermultiplets}

To describe operators from 1/2-BPS supermultiplets  in a manifestly supersymmetric and $SU(4)_R$ covariant way it is useful to consider the harmonic superspace parameterized by the set
of coordinates \cite{N=4_Harmonic_SS,SuperCor1}:
\begin{eqnarray}
\mbox{$\mathcal{N}=4$ harmonic
superspace}&=&\{x^{\alpha\dot{\alpha}},
~\theta^{+a}_{\alpha},\theta^{-a'}_{\alpha},
~\bar{\theta}_{+a~\dot{\alpha}},\bar{\theta}_{-a'~\dot{\alpha}},u
\}.
\end{eqnarray}
Here $u$ stands for a set of harmonic variables, parameterizing coset
$$
\frac{SU(4)}{SU(2) \times SU(2)' \times U(1)}
$$
and $a$ and $a'$ are the $SU(2)$ indices, $\pm$ corresponds
to $U(1)$ charge; $\theta$'s are Grassmann coordinates,
$\alpha$ and $\dot{\alpha}$ are the $SL(2,\mathbb{C})$ indices. More details on harmonic superspace and conventions used in this paper can be found in appendix \ref{aA}. We will also not write part of the indices explicitly in some expressions below, when it does not lead to confusion. On this superspace we can
define superfield $W^{++}(x,\theta^+,\bar{\theta}_-,u)$ which contains all fields of $\mathcal{N}=4$ SYM lagrangian, namely 6 $\phi^{AB}$ scalars (anti-symmetric in the $SU(4)_R$ indices $AB$), fermionic fields $\psi^A_{\alpha},\bar{\psi}^A_{\dot{\alpha}}$  and the gauge field strength tensor $F^{\mu\nu}$ , all transforming in the adjoint representation of the $SU(N_c)$ gauge group. $W^{++}$ is a constrained superfield in a sense, that its algebra of supersymmetry transformations for component fields  is closed only on their equations of motion. It is possible to consider so called chiral truncation of $W^{++}$  by putting $\bar{\theta}_-=0$ by hand: $W^{++}(x,\theta^+,0,u)$. In this case,  all component fields in $W^{++}(x,\theta^+,0,u)$ belong to self dual sector of the theory and their supersymmetry transformation could be closed off shell. In terms of component fields $W^{++}(x,\theta^+,0,u)$ is
written as:
\begin{eqnarray}
W^{++}(x,\theta^+,u)=\phi^{++}
+i\sqrt{2}\theta^{+a}_{\alpha}\epsilon_{ab}\epsilon^{\alpha\beta}\psi^{+b}_{\beta} + -i\frac{\sqrt{2}}{2}\theta^{+a}_{\alpha}\epsilon_{ab}\theta^{+b}_{\beta}F^{\alpha\beta} + ...,
\end{eqnarray}
where $\phi^{++}(x,u)=-1/2u^{+a}_A\epsilon_{ab}u^{+b}_B\phi^{AB}$,
$\psi_{\alpha}^{+a}(x,u) = u^{+a}_A\psi^A_{\alpha}$ and $F^{\alpha\beta}(x) = -\frac{1}{2}F_{\mu\nu}(\sigma^{\mu}\bar{\sigma}^{\nu})^{\alpha\beta}$. As usual, all components of $W^{++}$ superfield could be obtained from the lowest one ($\phi^{++}$) by the action of corresponding supercharges $Q_{\alpha}^{+a}$.

The 1/2-BPS supermultiplets of operators we wish to consider are a generalization of the chiral part of the stress-tensor supermultiplet $\mathcal{T}_2$. They are defined as
\begin{eqnarray}
\mathcal{T}_{p}=Tr([W^{++}(x,\theta^+,u)]^p).
\end{eqnarray}
With the help of supercharges $\mathcal{T}_p$ could be conveniently written as
\begin{eqnarray}\label{Wcondition1}
\mathcal{T}_p(x,\theta^+,u)=exp(\theta^{+a\alpha}Q_{+a\alpha})Tr([\phi^{++}\phi^{++}]^p).
\end{eqnarray}
Note, also, that the lowest components $\mathcal{T}_p(x,0,u)$ of operator supermultiplets are annihilated by half of the chiral and anti-chiral supercharges of the theory:
\begin{eqnarray}\label{Wcondition2}
[\mathcal{T}_p(x,0,u),Q_{-a'\alpha}]=0,
~[\mathcal{T}_p(x,0,u),\bar{Q}^{+a}_{\dot{\alpha}}]=0.
\end{eqnarray}

To describe on-shell states of the $\mathcal{N}=4$ supemultiplet we use on-shell momentum superspace introduced by Nair \cite{Nair}. Its ordinary and harmonic versions are parameterized by the following set of coordinates:
\begin{eqnarray}
\mbox{$\mathcal{N}=4$
on-shell momentum superspace}&=&\{\lambda_{\alpha},\tilde{\lambda}_{\dot{\alpha}},~\eta_A\},
\end{eqnarray}
or
\begin{eqnarray}
\mbox{$\mathcal{N}=4$ harmonic
on-shell momentum superspace}&=&\{\lambda_{\alpha},\tilde{\lambda}_{\dot{\alpha}},~\eta_{+a},\eta_{-a'},~u\}. \nonumber \\
\end{eqnarray}
Here $\lambda_{\alpha},\tilde{\lambda}_{\dot{\alpha}}$ are the $SL(2,\mathbb{C})$
commuting spinors that parameterize the
momentum carried by on-shell external state ( $p_{\alpha\dot{\alpha}}=\lambda_{\alpha}\tilde{\lambda}_{\dot{\alpha}}$ for $p^2=0$) and
$\eta_A$ or its harmonic projections $\eta_{+a},\eta_{-a'}$ are Grassmann coordinates. The latter are scalars with respect to Lorentz transformations.

All creation/annihilation operators of on-shell states, which are
two physical polarizations of gluons $|g^-\rangle, |g^+\rangle$,
four fermions $|\Gamma^A\rangle$ with positive and four fermions
$|\bar{\Gamma}_A\rangle$ with negative helicity together with three complex scalars $|\phi^{AB}\rangle$ (anti-symmetric in the $SU(4)_R$ indices
$AB$ ) can be combined together into one $\mathcal{N}=4$ invariant
superstate ("superwave-function")
$|\Omega_{i}\rangle=\Omega_{i}|0\rangle$ ($i$ labels particular external state):
\begin{eqnarray}\label{superstate}
|\Omega_{i}\rangle=\left(g^+_i + \eta_A\Gamma^{A}_{i} +
\frac{1}{2!}\eta_A\eta_B\phi_{i}^{AB} +
\frac{1}{3!}\eta_A\eta_B\eta_C\varepsilon^{ABCD}\bar{\Gamma}_{i,D} +
\frac{1}{4!}\eta_A\eta_B\eta_C\eta_D\varepsilon^{ABCD} g^-_i\right)
|0\rangle, \nonumber \\
\end{eqnarray}
where $\varepsilon^{ABCD}$ is Levi-Civita symbol. The $n$ particle superstate
$|\Omega_n\rangle$ is then given by
$|\Omega_n\rangle=\prod_{i=1}^n\Omega_i|0\rangle$.
Note, that on-shell momentum superspace is chiral.

We can then formally write down the form factors $\mathcal{F}_{\bold{p},n}$ of 1/2-BPS supermultiplets $\mathcal{T}_p$ introduced above as:
\begin{eqnarray}
\mathcal{F}_{\bold{p},n}(\{\lambda,\tilde{\lambda},\eta\},x,\theta^{+})=
\langle\Omega_n|\mathcal{T}_p(x,\theta^{+})|0\rangle,
\end{eqnarray}
Here we are considering the color
ordered object $\mathcal{F}_{\bold{p},n}$. The physical form factor $\mathcal{F}_{\bold{p},n}^{phys}$ in the planar limit\footnote{$g \rightarrow 0$ and $N_c \rightarrow \infty$ of $SU(N_c)$ gauge group so that $\lambda=g^2N_c=$fixed.}  should be obtained from $\mathcal{F}_{\bold{p},n}$ as usual.
\begin{eqnarray}
\mathcal{F}_{\bold{p},n}^{phys.}(\{\lambda,\tilde{\lambda},\eta\},x,\theta^{+})=
(2\pi)^4g^{n-2}2^{n/2}\sum_{\sigma\in
S_n/Z_n}Tr(t^{a_{\sigma(1)}}\ldots
t^{a_{\sigma(n)}})\mathcal{F}_{\bold{p},n}(\sigma(\{\lambda,\tilde{\lambda},\eta\}),x,\theta^{+}), \nonumber \\
\end{eqnarray}
where the sum runs over all possible none-cyclic permutations
$\sigma$ of the set $\{\lambda,\tilde{\lambda},\eta\}$ and the trace
involves $SU(N_c)$ $t^a$ generators in the fundamental
representation. The normalization $Tr(t^at^b)=1/2$ is used.

The object, which we will actually analyze is the super Fourier transform of the coordinate superspace form factor:
\begin{equation}
\hat{T}[\ldots] = \int d^4x~d^{-4}\theta
\exp (-iqx+i\theta^{+a\alpha}\gamma_{+a\alpha})[\ldots],
\end{equation}
\begin{equation}\label{T[superFormfactor]}
Z_{\bold{p},n} (\{\lambda,\tilde{\lambda},\eta\},\{q,\gamma_{+}\}) =
\hat{T}[\mathcal{F}_{\bold{p},n}].
\end{equation}
Note that while $\mathcal{F}_{\bold{p},n}$ carries $+2p$ $U(1)$ charge, after
$\hat{T}$ transformation it will be reduced to $+2p-4$ for $Z_{\bold{p},n}$. Taking into account that $\mathcal{F}_{\bold{p},n}$ is chiral and translationally
invariant, while $\mathcal{T}_p$ is $1/2$-BPS and considering the corresponding Ward identities, we see that the form factor
$\mathcal{F}_{\bold{p},n}$ should satisfy the following set of conditions
\cite{BKV_SuperForm}:
\begin{eqnarray}
P_{\alpha\dot{\alpha}}\mathcal{F}_{\bold{p},n}=Q_{-a\alpha}\mathcal{F}_{\bold{p},n}=Q_{-a'\alpha}\mathcal{F}_{\bold{p},n}=\bar{Q}^{+a\dot{\alpha}}\mathcal{F}_{\bold{p},n}=0,
\end{eqnarray}
where generators of supersymmetry algebra
$(P_{\alpha\dot{\alpha}},Q_{+a\alpha},Q_{-a'\alpha},\bar{Q}^{+a}_{\dot{\alpha}},\bar{Q}^{-a'}_{\dot{\alpha}})$
acting on $\mathcal{F}_{\bold{p},n}$ are given by
\begin{eqnarray}\label{SUSYGenF}
\mbox{4 translations } P_{\alpha\dot{\alpha}}&=&
-\sum_{i=1}^n\lambda_{\alpha,i}\tilde{\lambda}_{\dot{\alpha},i}+q_{\alpha\dot{\alpha}},\nonumber\\
\mbox{4 supercharges }
Q_{+a\alpha} &=&-\sum_{i=1}^n\lambda_{\alpha,i}\eta_{+a,i}
+\frac{\partial}{\partial\theta^{+a\alpha} },\nonumber\\
\mbox{4 supercharges }
Q_{-a'\alpha}&=&-\sum_{i=1}^n\lambda_{\alpha,i}\eta_{-a',i}
+ \frac{\partial}{\partial\theta^{-a'\alpha} },\nonumber\\
\mbox{4 conjugated supercharges }
\bar{Q}^{+a}_{\dot{\alpha}}&=&-\sum_{i=1}^n\tilde{\lambda}_{\dot{\alpha},i}
\frac{\partial}{\partial\eta_{+a,i}}+\theta^{+a\alpha}q_{\alpha\dot{\alpha}},\nonumber\\
\mbox{4 conjugated supercharges }
\bar{Q}^{-a'}_{\dot{\alpha}}&=&-\sum_{i=1}^n\tilde{\lambda}_{\dot{\alpha},i}
\frac{\partial}{\partial\eta_{-a',i}}+\theta^{-a'\alpha}q_{\alpha\dot{\alpha}}.
\end{eqnarray}
These relations imply, that $\mathcal{F}_{\bold{p},n}$ takes the following form \cite{BKV_SuperForm}
\begin{eqnarray}\label{T[superFormfactor]}
Z_{\bold{p},n} (\{\lambda,\tilde{\lambda},\eta\},\{q,\gamma_{+}\}) &=&
\delta^4(\sum_{i=1}^n\lambda_{\alpha,i}\tilde{\lambda}_{\dot{\alpha},i}-q_{\alpha\dot{\alpha}})
\delta^{-4}(q_{+a\alpha}+\gamma_{+a\alpha})\delta^{+4}(q_{-a'\alpha})
\mathcal{X}_{p,n}\left(\{\lambda,\tilde{\lambda},\eta\}\right),\nonumber\\
\mathcal{X}_{p,n}&=&\mathcal{Y}_{p,n}^{(2p-4)}+ \mathcal{Y}_{p,n}^{(2p)} + \ldots +
\mathcal{Y}_{p,n}^{(4n+2p-12)}.
\end{eqnarray}
$\mathcal{Y}^{m}_{p,n}$ are the homogenous $SU(4)_R$ and $SU(2)\times SU(2)'$ invariant polynomials in Grassmann variables of order $m$ carrying $2p-4$ units of $U(1)$ charge. Here
\begin{eqnarray}
q_{+a\alpha}=\sum_{i=1}^n\lambda_{\alpha,i}\eta_{+a,i},
~q_{-a'\alpha}=\sum_{i=1}^n\lambda_{\alpha,i}\eta_{-a',i}.
\end{eqnarray}
Grassmann delta functions are defined as
(see the appendix A for the whole set of definitions regarding Grassmann delta functions and their integration)
\begin{eqnarray}
\delta^{-4}\left(q_{+a\alpha}\right)&=&\prod_{a,b=1}^2
\epsilon^{\alpha\beta}q_{+a,\alpha}q_{+b\beta},
\nonumber\\
\delta^{+4}\left(q_{-a'\alpha}\right)&=&\prod_{a',b'=1}^2
\epsilon^{\alpha\beta}q_{-a'\alpha}q_{-b'\beta}.
\end{eqnarray}
To save space we also use the notation:
\begin{eqnarray}
\delta^{8}(q+\gamma)\equiv\delta^{-4}(q_+ + \gamma_+)\delta^{+4}(q_-)\equiv
\delta^{-4}(q_{+a\alpha}+\gamma_{+a\alpha})\delta^{+4}(q_{-a'\alpha}).
\end{eqnarray}
We will also drop momentum conservation delta function where it will not
lead to confusion.

Note, that $\mathcal{Y}^{(2p-4)}_n$, $\mathcal{Y}^{(2p)}_n$ etc. in (\ref{T[superFormfactor]}) are understood as
analogs \cite{DualConfInvForAmplitudesCorch} of the MHV, NMHV etc.
parts of the superamplitude. For example, at tree level using BCFW recursion it is easy to obtain, that in the case of $p=2$ (stress tensor supermultiplet):
\begin{eqnarray}
\mathcal{Y}^{(0)}_{2,n}=\mathcal{X}^{(0)}_n,
~\mathcal{X}^{(0)}_n=\frac{1}{\langle12\rangle\langle23\rangle...\langle n1\rangle},
\end{eqnarray}
and in the case of $p=3$ we have \cite{General1/2BPSformFactors}
\begin{eqnarray}
\mathcal{Y}^{(2)}_{3,n}=\mathcal{X}^{(0)}_n\sum_{i<j=1}^n \langle ij\rangle\frac{1}{2}\eta_{+a,i}\epsilon^{ab}\eta_{+b,j}.
\end{eqnarray}
Also, for completeness let us write down well known answers for tree level
$\mbox{MHV}_n$ and $\overline{\mbox{MHV}}_3$ amplitudes (the total momentum conservation delta function is dropped)
\begin{equation}
A_n^{(0)MHV}=\frac{\delta^{8}(q)}{\langle 12\rangle\langle23\rangle...\langle
n1\rangle},~
A_3^{(0)\overline{MHV}}=\frac{\hat{\delta}^{4}(\eta_1[23]+\eta_2[31]+\eta_3[12])}{[12][23][31]},
\end{equation}
which we will use throughout this paper.

\subsection{Form factors of Konishi operator supermultiplet}

At present there is no manifestly supersymmetric and $SU(4)_R$ covariant formulation of operators from Konishi supermultiplet similar to the case of 1/2-BPS supermultiplets considered above. However, we may proceed considering form factors of the lowest component
of Konishi supermultiplet, where only external states are taken into account in manifestly supersymmetric way \cite{Wilhelm_AmplitudesFormFactorsDilatationOperator}. The lowest component of Konishi supermultiplet is given by operator
\begin{equation}
\mathcal{K}=\frac{1}{8}\epsilon^{ABCD}Tr(\phi_{AB}\phi_{CD}).
\end{equation}
Using ordinary on shell momentum superspace  the color ordered
form factors of $\mathcal{K}$ could be written as:
\begin{equation}
Z_{\mathcal{K},n}(\{\lambda,\tilde{\lambda},\eta\},q)=
\langle\Omega_n|\mathcal{K}(q)|0\rangle.
\end{equation}
At tree level we have:
\begin{equation}
Z_{\mathcal{K},2}=
\delta^4(\sum_{i=1}^2\lambda_{\alpha,i}\tilde{\lambda}_{\dot{\alpha},i}-q_{\alpha\dot{\alpha}})
\left(\varepsilon^{ABCD}(\eta_{A,1}\eta_{B,1})(\eta_{C,2}\eta_{D,2})\right),
\end{equation}
and
\begin{eqnarray}
Z_{\mathcal{K},3}&=&
\delta^4(\sum_{i=1}^3\lambda_{\alpha,i}\tilde{\lambda}_{\dot{\alpha},i}-q_{\alpha\dot{\alpha}})
\frac{\left(1+\mathbb{P}+\mathbb{P}^2\right)}{\langle12\rangle\langle23\rangle\langle 31\rangle}\times\nonumber\\
&\times&\left(\langle12\rangle^2\varepsilon^{ABCD}(\eta_{A,1}\eta_{B,1})(\eta_{C,2}\eta_{D,2})
+2\langle13\rangle\langle23\rangle
\varepsilon^{ABCD}(\eta_{A,1}\eta_{B,2})(\eta_{C,3}\eta_{D,3})\right),\nonumber\\
\end{eqnarray}
where $\mathbb{P}$ is permutation operator which permutes indices of external states, i.e. for example,
$\mathbb{P}(\langle13\rangle\eta_{A,1}\eta_{C,3})=\langle21\rangle\eta_{A,2}\eta_{C,1}$ for $n=3$.

\section{Soft theorems for form factors in $\mathcal{N}=4$ SYM}\label{p3}

In this chapter we are going first to briefly remind you essential details of the derivation of soft theorems for the case of gluon amplitudes \cite{Casali_SoftSubLeadingYM} and then proof similar statements for the case of 1/2-BPS form factors at tree level. Also we will verify validity of soft theorems on some particular examples in the case of stress tensor supermultiplet and Konishi supermultiplet form factors. Next, we going to consider one-loop corrections to the soft theorems in the case of $n=3,4$ form factors of stress tensor supermultiplet
and in the case of $n=3$ Konishi form factors.

\begin{figure}[t]
 \begin{center}
 \leavevmode
  \epsfxsize=6cm
 \epsffile{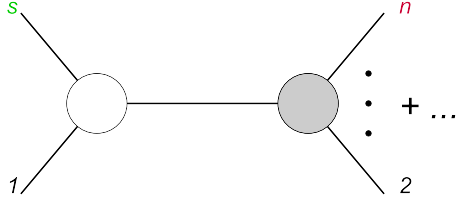}
 \end{center}\vspace{-0.2cm}
 \caption{Singular term in BCFW recursion. The $[\textcolor{green}{s},\textcolor{red}{n} \rangle$ shift. White blob is $\overline{\mbox{MHV}}_3$ amplitude.}\label{BCFWamplitude}
 \end{figure}

So, consider  tree level color ordered pure gluon amplitude\footnote{We will omit total momentum conservation delta function throughout this chapter.}
$$
A_{n+1}^{h_s,h_1,...,h_n}(p_s,p_1,...,p_n)
$$
with on-shell external particles having momenta $p_s,p_1,...,p_n$ and helicities $h_s=+,h_1,...,h_n$ written in terms of spinor helicity variables ($p_i=\lambda_i\tilde{\lambda}_i$):
$$
A_{n+1}^{h_s,h_1,...,h_n}(\{\lambda_s,\tilde{\lambda}_s\},\{\lambda_1,\tilde{\lambda}_1\},...,\{\lambda_n,\tilde{\lambda}_n\}).
$$
The general idea behind the proof of the soft theorem in the amplitude case is to consider BCFW representation \cite{BCFW1,BCFW2} of $n+1$ point amplitude $A_{n+1}^{+,h_1,...,h_n}$ with  $[n,s\rangle$ shift
\begin{eqnarray}
&&\hat{\lambda}_{s}=\lambda_{s}+z\lambda_{n},\nonumber\\
&&\hat{\tilde{\lambda}}_{n}=\tilde{\lambda}_{n}-z\tilde{\lambda}_{s}.
\end{eqnarray}
The amplitude in this case is given by a sum of products of lower point amplitudes (see Fig. \ref{BCFWamplitude})
\begin{eqnarray}
&&A_{n+1}^{+,h_1,...,h_n}(\{\lambda_s,\tilde{\lambda}_s\},\{\lambda_1,\tilde{\lambda}_1\},...,\{\lambda_n,\tilde{\lambda}_n\})=\sum_{h_I=\pm}\sum_{i=1}^{n-2}\frac{1}{s_{s...i}}\times
\nonumber\\
&\times&
A_L^{+,h_1,...,h_i,h_I}(\{\hat{\lambda}_s,\tilde{\lambda}_s\},...,\{\lambda_i,\tilde{\lambda}_i\},\{\hat{\lambda}_I,\hat{\tilde{\lambda}}_I\})
A_R^{-h_I,h_{i+1},...,h_n}(\{\hat{\lambda}_I,-\hat{\tilde{\lambda}}_I\},\{\lambda_{i+1},\tilde{\lambda}_{i+1}\},...,\{\lambda_n,\hat{\tilde{\lambda}}_n\}),\nonumber\\
\end{eqnarray}
together with standard BCFW substitutions:
\begin{eqnarray}
\hat{p}_I&=&\lambda_i\tilde{\lambda}_i+...+(\lambda_{s}+z_i\lambda_{n})\tilde{\lambda}_s,\nonumber\\
\hat{p}_s&=&(\lambda_{s}+z_i\lambda_{n})\tilde{\lambda}_s,\nonumber\\
\hat{p}_n&=&\lambda_n(\tilde{\lambda}_{n}-z_i\tilde{\lambda}_{s}),
\nonumber\\
z_i&=&-\frac{s_{s,...,i}}{\langle n|s+...+i|s]}.
\end{eqnarray}
Here, standard $s_{i...j}\equiv(p_i+...p_j)^2$ notation was used.
Note also that due to the little group scaling properties if $h_i=+$
\begin{equation}
A_n^{h_1,...,h_{i-1},+,h_{i+1},...,h_n}(...,\{\sqrt{\epsilon}\lambda_i,\sqrt{\epsilon}\tilde{\lambda}_i\},...)
=\epsilon A_n^{h_1,...,h_{i-1},+,h_{i+1},...,h_n}(...,\{\epsilon\lambda_i,\tilde{\lambda}_i\},...),
\end{equation}
So, to study soft behavior with respect to $p_s$ momentum we may consider pure holomorphic rescaling:  $\lambda_s\mapsto \epsilon \lambda_s$ only, and analyze the poles in $1/\epsilon^2$ and $1/\epsilon$ in $\epsilon$ as $\epsilon \rightharpoonup 0$.
It is easy to see, that in the limit $p_s \rightarrow \epsilon p_s,~ \epsilon \rightarrow 0$ the singularities in $\epsilon$ will come only from the term $A^{\overline{MHV}}\otimes A_n$ in BCFW recursion\footnote{$\otimes$ stands for the summation over internal states and substitution of the corresponding $z$ values.} (in this case $h_I=-h_1$ and the result is the same both for $h_1=+$ and $h_1=-$; thus, in what follows we may choose $h_1=+$):
\begin{eqnarray}
&&A_{n+1}^{+,h_1,...,h_n}(\{\epsilon\lambda_s,\tilde{\lambda}_s\},\{\lambda_1,\tilde{\lambda}_1\},...\{\lambda_n,\tilde{\lambda}_n\})=
A_3^{++-}(\{\epsilon\hat{\lambda}_s,\tilde{\lambda}_s\},\{\lambda_1,\tilde{\lambda}_1\},\{\hat{\lambda}_I,\hat{\tilde{\lambda}}_I\})\times\nonumber\\
&\times&
\frac{1}{s_{s1}}A_n^{h_1,...,h_n}(\{\hat{\lambda}_I,-\hat{\tilde{\lambda}}_I\},\{\lambda_1,\tilde{\lambda}_1\},...,\{\lambda_n,\hat{\tilde{\lambda}}_n\})+O(\epsilon).
\end{eqnarray}
Using explicit expression for $A_3^{\overline{MHV}}$ amplitude
\begin{eqnarray}
A_3^{++-}(\{\lambda_1,\tilde{\lambda}_1\},\{\lambda_2,\tilde{\lambda}_2\},\{\lambda_3,\tilde{\lambda}_3\})
=\frac{[12]^4}{[12][23][31]},
\end{eqnarray}
together with explicit spinor expressions for the internal state
\begin{eqnarray}
&& \hat{\lambda}_I = \lambda_1, \nonumber \\
&& \hat{\tilde{\lambda}}_I = -\tilde{\lambda}_1 - \epsilon\frac{\langle n s\rangle}{\langle n 1 \rangle}\tilde{\lambda}_s,
\end{eqnarray}
and expanding the result in powers of $\epsilon$ we have
\begin{eqnarray}
&&A_{n+1}^{+,h_1,...,h_n}(\{\epsilon\lambda_s,\tilde{\lambda}_s\},\{\lambda_1,\tilde{\lambda}_1\},...,\{\lambda_n,\tilde{\lambda}_n\})=
\nonumber\\
&=&\left(\frac{\hat{S}_1}{\epsilon^2}+\frac{\hat{S}_2}{\epsilon}\right)A_{n}^{h_1,...,h_n}(\{\lambda_1,\tilde{\lambda}_1\},...\{\lambda_n,\tilde{\lambda}_n\})
+O(\epsilon),
\end{eqnarray}
with
\begin{eqnarray}
\hat{S}_1=\frac{\langle 1n\rangle}{\langle ns\rangle\langle s1\rangle },
~\hat{S}_2=\frac{\tilde{\lambda}^{\dot{\alpha}}_s}{\langle s1 \rangle}\frac{\partial}{\partial\tilde{\lambda}^{\dot{\alpha}}_1}+
\frac{\tilde{\lambda}^{\dot{\alpha}}_s}{\langle sn \rangle}\frac{\partial}{\partial\tilde{\lambda}^{\dot{\alpha}}_n}.
\end{eqnarray}

\subsection{Tree level $\mathcal{N}=4$ SYM form factors}

Now let's consider BCFW recursion for tree level 1/2-BPS form factors $\mathcal{T}_{2,n}$ (stress tensor operator supermultiplet)
\begin{eqnarray}
Z_{\bold{2},n+1}^{(0)}(\{\lambda_s,\tilde{\lambda}_s,\eta_s\}, \{\lambda_1,\tilde{\lambda}_1,\eta_1\},...,\{\lambda_n,\tilde{\lambda}_n,\eta_n\};q,\gamma)
\end{eqnarray}
In what follows, we will drop $(0)$ superscript everywhere below, where it will not lead to confusion.  Using $[n,s\rangle$ supersymmetric BCFW shift,
\begin{eqnarray}
&&\hat{\lambda}_{s}=\lambda_{s}+z\lambda_{n},\nonumber\\
&&\hat{\tilde{\lambda}}_{n}=\tilde{\lambda}_{n}-z\tilde{\lambda}_{s},\nonumber\\
&&\hat{\eta}_{n}=\eta_{n}+z\eta_{s},
\end{eqnarray}
having correct large $z$ behavior\footnote{There is no need to worry about boundary terms \cite{General1/2BPSformFactors}.}, we get
\begin{eqnarray}\label{StressTensorBCFWGeneral}
Z_{\bold{2},n+1}&=&\sum_{i=1}^{n-2} \int d^4\eta_I
A_{i+1}(\{\hat{\lambda}_s, \hat{\tilde{\lambda}}_s, \eta_s\},\ldots , \{\lambda_i, \tilde{\lambda}_i, \eta_i\}, \{\hat{\lambda}_I, \hat{\tilde{\lambda}}_I, \hat{\eta}_I\})\times \nonumber \\
&\times & \frac{1}{s_{s,1\ldots i}} Z_{\bold{2},n-i+1}(\{\hat{\lambda}_I,-\hat{\tilde{\lambda}}_I, \hat{\eta}_I\}, \{\lambda_{i+1}, \tilde{\lambda}_{i+1}, \eta_{i+1}\}, \ldots ,\{\hat{\lambda}_n,\hat{\tilde{\lambda}}_n,\hat{\eta}_n\}; q,\gamma) \nonumber \\
&+& \sum_{i=s,1}^{n-2}\int d^4\eta_I Z_{\bold{2},i+1} (\{\hat{\lambda}_s, \hat{\tilde{\lambda}}_s, \eta_s\}, \ldots , \{\lambda_i, \tilde{\lambda}_i, \eta_i\}, \{\hat{\lambda}_I, \hat{\tilde{\lambda}}_I,\hat{\eta}_I\}; q,\gamma)\times \nonumber \\
&\times & \frac{1}{s_{s,1\ldots i}} A_{n-i+1} (\{\hat{\lambda}_I,-\hat{\tilde{\lambda}}_I, \hat{\eta}_I\}, \{\lambda_{i+1},\tilde{\lambda}_{i+1},\eta_{i+1}\},\ldots , \{\hat{\lambda}_n,\hat{\tilde{\lambda}}_n, \hat{\eta}_n\}),
\end{eqnarray}
where $s_{s,1...i}=(p_s+p_1+...+p_i)^2$ and the subscript $i=s,1$ in the sum above is understood  in a sense, that if $i=s$ then $i+1\mapsto1$.

Here, for super form factors we expect the same rescaling properties under little group transformations of helicity spinors
as in the case of amplitudes and thus will consider holomorphic rescaling $\lambda_s \mapsto \epsilon \lambda_s$ when taking soft limit.

In a case when soft leg belongs to amplitude it is easy to see, that
the only divergent contribution in the limit $\epsilon \rightarrow 0$ for $Z_{\bold{p},n+1}$ will come from the term with $i=1$ in first line of (\ref{StressTensorBCFWGeneral}), which is given by $(A_3^{\overline{MHV}} \otimes Z_{\bold{2},n})$, while all other terms are regular in this limit. This term is given by
\begin{eqnarray}
\int d^4\eta_I \frac{\hat{\delta}^4(\eta_I[s1]+\eta_s[1\hat{I}]+\eta_1[\hat{I}s])}{[s1][1\hat{I}][\hat{I}s]}
~\frac{1}{s_{s1}}~Z_{\bold{2},n}(\{\hat{\lambda}_I,-\hat{\tilde{\lambda}}_I,\eta_I\},
\{\lambda_2,\tilde{\lambda}_2,\eta_2\},...,\{\hat{\lambda}_{n},\hat{\tilde{\lambda}}_n,\hat{\eta}_n\};q,\gamma).
\nonumber\\
\end{eqnarray}
Performing Grassmann integration and substituting corresponding $z$ value we get
\begin{eqnarray}
&&\frac{[s1]^3}{[\hat{I}1][\hat{I}s][s1]\langle1s\rangle}\times\nonumber\\&&\times Z_{\bold{2},n}\left(
\{\lambda_1,\tilde{\lambda}_1+\frac{\langle ns \rangle}{\langle n1 \rangle}\tilde{\lambda}_s,\eta_1+\frac{\langle ns \rangle}{\langle n1 \rangle}\eta_s\},
\{\lambda_2,\tilde{\lambda}_2,\eta_2\},...,\{\lambda_{n},\tilde{\lambda}_n-\frac{\langle 1s \rangle}{\langle n1 \rangle}\tilde{\lambda}_s,\eta_n-\frac{\langle 1s \rangle}{\langle n1 \rangle}\eta_s\};q,\gamma\right).
\nonumber
\end{eqnarray}
Coefficient in front of $Z_{\bold{2},n}$ can be simplified as
\begin{eqnarray}
\frac{[s1]^3}{[\hat{I}1][\hat{I}s][s1]\langle1s\rangle}=\frac{\langle n1\rangle}{\langle ns \rangle \langle s1\rangle}.
\end{eqnarray}
Rescaling $\lambda_s \mapsto \epsilon\lambda_s$, $\tilde{\lambda}_s \mapsto \tilde{\lambda}_s$, $\eta_s \mapsto \eta_s$ and
performing Taylor expansion of $Z_{\bold{2},n}$ up to the order $O(\epsilon^2)$ we get:
\begin{eqnarray}
\left(\frac{\hat{S}_1}{\epsilon^2}+\frac{\hat{S}_2}{\epsilon}\right)
Z_{\bold{2},n}\left({\{\lambda_1,\tilde{\lambda}_1,\eta_1\},...,\{\lambda_n,\tilde{\lambda}_n,\eta_n\}};q,\gamma\right)
+reg.,~\epsilon \rightarrow0
\end{eqnarray}
where
\begin{eqnarray}
\hat{S}_1=\frac{\langle 1n\rangle}{\langle ns\rangle\langle s1\rangle },
~\hat{S}_2=\frac{\tilde{\lambda}^{\dot{\alpha}}_s}{\langle s1 \rangle}\frac{\partial}{\partial\tilde{\lambda}^{\dot{\alpha}}_1}+
\frac{\tilde{\lambda}^{\dot{\alpha}}_s}{\langle sn \rangle}\frac{\partial}{\partial\tilde{\lambda}^{\dot{\alpha}}_n}+
\frac{\eta_{\Lambda,s}}{\langle s1 \rangle}\frac{\partial}{\partial \eta_{\Lambda,1}}+
\frac{\eta_{\Lambda,s}}{\langle sn \rangle}\frac{\partial}{\partial \eta_{\Lambda,n}}.
\end{eqnarray}
Here $\Lambda$ is $SU(4)_R$ index which combines $SU(2)\times SU(2)'\times U(1)$ indices $+a$ and $-a'$.
\begin{figure}[t]
 \begin{center}
 \leavevmode
  \epsfxsize=14cm
 \epsffile{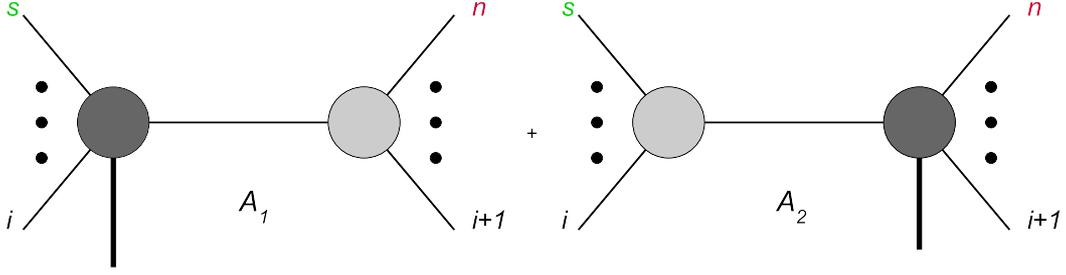}
 \end{center}\vspace{-0.2cm}
 \caption{Schematic representation of BCFW recursion for form factors. The $[\textcolor{green}{s},\textcolor{red}{n} \rangle$ shift. Dark grey blob is form factor.}\label{2}
 \end{figure}
Other terms in BCFW recursion for form factors without 3-point amplitude are finite in the limit $\epsilon \rightarrow0$ for the same reason as in the case of amplitudes. In all these other terms (this can be seen directly from BCFW substitutions) in the limit $\epsilon \mapsto 0$ after rescaling we have $\hat{\lambda}_s \mapsto fin.$ and $\hat{\tilde{\lambda}}_s=\tilde{\lambda}_s \mapsto fin.$
On the contrary, in the case of a term with 3-point amplitude we have $\hat{\lambda}_s \sim \epsilon$ after rescaling. This is exactly the source of singular behavior of the form factor or amplitude in the soft limit $\epsilon \rightarrow0$. The case, when a soft leg belongs to the  form factor, is no different from the case when soft leg
belongs to the amplitude. In general, this contribution is finite.  The only special case is given by a term with 2-point form factor (these terms give contributions which have no direct analogs in the amplitude case) and it is in fact explicitly finite: $Z_{\bold{2},2}$ does not contain negative powers of spinors at all:
\begin{eqnarray}
Z_{\bold{2},2}&=&\frac{\delta^8(\lambda_1\eta_1+\lambda_2\eta_2+\gamma)}{\langle12\rangle^2}=
\frac{\delta^{-4}(\lambda_1\eta_1+\lambda_2\eta_2+\gamma)\delta^{+4}(\lambda_1\eta_1+\lambda_2\eta_2)}{\langle12\rangle^2}
\nonumber\\&=&
\delta^{-4}(\lambda_1\eta_1+\lambda_2\eta_2+\gamma)\eta_{-1,1}\eta_{-2,1}\eta_{-1,2}\eta_{-2,2}.
\end{eqnarray}
The factors $1/(p_s+q)^2$ and $z=-(q+p_s)^2/\langle n|q|s]$ are also finite in the limit $p_s \rightarrow 0$.
So finally we can write, that
\begin{eqnarray}
&&Z_{\bold{2},n+1}\left(\{\epsilon\lambda_s,\tilde{\lambda}_s,\eta_s\},{\{\lambda_1,\tilde{\lambda}_1,\eta_1\},...,\{\lambda_n,\tilde{\lambda}_n,\eta_n\}};q,\gamma\right)=
\nonumber\\
&&\left(\frac{\hat{S}_1}{\epsilon^2}+\frac{\hat{S}_2}{\epsilon}\right)
Z_{\bold{2},n}\left({\{\lambda_1,\tilde{\lambda}_1,\eta_1\},...,\{\lambda_n,\tilde{\lambda}_n,\eta_n\}};q,\gamma\right)
+reg.,~\epsilon \rightarrow0
\end{eqnarray}
It is also interesting to note that in the limit when (super)momentum carried by operator goes to zero $(q,\gamma) \mapsto 0$ form factors have different, but still universal and well defined behavior (see \cite{BKV_SuperForm}):
\begin{equation}\label{cojectureAmpl-FF}
Z_{\bold{2},n}(\{\lambda,\tilde{\lambda},\eta\};0,0)= g\frac{\partial
A_{n}(\{\lambda,\tilde{\lambda},\eta\})}{\partial g}.
\end{equation}
Note, that this relation should be valid not only at tree level but to all orders in loop expansion.

Now let's turn to the general 1/2-BPS ($p>2$) form factors \cite{General1/2BPSformFactors}. It is not known much about BCFW recursion for such form factors. The problem is with non-vanishing behavior of these form factors in large $z$ limit for general BCFW shifts.  However, in the case of $p=3$ it is known, that next to adjacent BCFW shift works fine (form factor vanishes in large $z$ limit) for general $\mbox{N}^k\mbox{MHV}$ form factors. So, consider the super form factor
\begin{eqnarray}
Z_{\bold{3},n+1}\left({\{\lambda_1,\tilde{\lambda}_1,\eta_1\},\{\lambda_s,\tilde{\lambda}_s,\eta_s\},\{\lambda_3,\tilde{\lambda}_3,\eta_3\},
...,\{\lambda_n,\tilde{\lambda}_n,\eta_n\},\{\lambda_{n+1},\tilde{\lambda}_{n+1},\eta_{n+1}\}};q,\gamma\right)\nonumber\\
\end{eqnarray}
and its decomposition via next to adjacent $[n+1,s\rangle$ shift:
\begin{eqnarray}
&&\hat{\lambda}_{s}=\lambda_{s}+z\lambda_{n+1},\nonumber\\
&&\hat{\tilde{\lambda}}_{n+1}=\tilde{\lambda}_{n+1}-z\tilde{\lambda}_{s},\nonumber\\
&&\hat{\eta}_{\pm,n+1}=\eta_{\pm,n+1}+z\eta_{\pm,s}.
\end{eqnarray}
We are interested in the soft behavior in the limit $\lambda_s \mapsto \epsilon \lambda_s$,
$\epsilon\rightarrow 0$. Just as in the discussion above, poles in $1/\epsilon$ come from contributions in BCFW decomposition, which involve $\mbox{MHV}_3$ and $\overline{\mbox{MHV}}_3$ amplitudes due to their degenerate kinematics. All contributions containing $\mbox{MHV}_3$ amplitudes equal to 0 for this particular shift.
So, once again we have to consider contributions with  $\overline{\mbox{MHV}}_3$ amplitudes only, but now we have two such contributions $A_1$ and $A_2$:
\begin{eqnarray}
A_1&=&\int d^4\eta_I \frac{\hat{\delta}^4(\eta_I[s1]+\eta_s[1\hat{I}]+\eta_1[\hat{I}s])}{[s1][1\hat{I}][\hat{I}s]}
~\frac{1}{s_{1s}}\times~
\nonumber\\&\times&Z_{\bold{3},n}(\{\hat{\lambda}_I,\hat{\tilde{\lambda}}_I,\eta_I\},
\{\lambda_3,\tilde{\lambda}_3,\eta_3\},...,\{\lambda_{n+1},\hat{\tilde{\lambda}}_{n+1},\hat{\eta}_{n+1}\};q,\gamma).
\nonumber\\
\end{eqnarray}
and
\begin{eqnarray}
A_2&=&\int d^4\eta_I \frac{\hat{\delta}^4(\eta_I[s3]+\eta_s[3\hat{I}]+\eta_3[\hat{I}s])}{[s3][3\hat{I}][\hat{I}s]}
~\frac{1}{s_{3s}}\times~
\nonumber\\&\times&Z_{\bold{3},n}(\{\lambda_1,\tilde{\lambda}_1,\eta_1\},\{\hat{\lambda}_I,\hat{\tilde{\lambda}}_I,\eta_I\},...,\{\lambda_{n+1},\hat{\tilde{\lambda}}_{n+1},\hat{\eta}_{n+1}\};q,\gamma).
\nonumber\\
\end{eqnarray}
Similar to the previous discussion, each term in the soft limit behaves as
\begin{eqnarray}
A_1=\left(\frac{\hat{S}_1^{(A_1)}}{\epsilon^2}+\frac{\hat{S}_2^{(A_1)}}{\epsilon}\right)
Z_{\bold{3},n}\left({\{\lambda_1,\tilde{\lambda}_1,\eta_1\},...,\{\lambda_n,\tilde{\lambda}_n,\eta_n\}};q,\gamma\right)
+reg.,~\epsilon \rightarrow0
\end{eqnarray}
\begin{eqnarray}
\hat{S}_1^{(A_1)}=\frac{\langle 1n+1\rangle}{\langle n+1s\rangle\langle s1\rangle },
~\hat{S}_2^{(A_1)}=\frac{\tilde{\lambda}^{\dot{\alpha}}_s}{\langle s1 \rangle}\frac{\partial}{\partial\tilde{\lambda}^{\dot{\alpha}}_1}+
\frac{\tilde{\lambda}^{\dot{\alpha}}_s}{\langle sn+1 \rangle}\frac{\partial}{\partial\tilde{\lambda}^{\dot{\alpha}}_{n+1}}+
\frac{\eta_{\Lambda,s}}{\langle s1 \rangle}\frac{\partial}{\partial \eta_{\Lambda,1}}+
\frac{\eta_{\Lambda,s}}{\langle sn+1 \rangle}\frac{\partial}{\partial \eta_{\Lambda,n+1}}.\nonumber\\
\end{eqnarray}
and
\begin{eqnarray}
A_2=\left(\frac{\hat{S}_1^{(A_2)}}{\epsilon^2}+\frac{\hat{S}_2^{(A_2)}}{\epsilon}\right)
Z_{\bold{3},n}\left({\{\lambda_1,\tilde{\lambda}_1,\eta_1\},...,\{\lambda_n,\tilde{\lambda}_n,\eta_n\}};q,\gamma\right)
+reg.,~\epsilon \rightarrow0
\end{eqnarray}
\begin{eqnarray}
\hat{S}_1^{(A_2)}=\frac{\langle 3n+1\rangle}{\langle n+1s\rangle\langle s3\rangle },
~\hat{S}_2^{(A_2)}=\frac{\tilde{\lambda}^{\dot{\alpha}}_s}{\langle s3 \rangle}\frac{\partial}{\partial\tilde{\lambda}^{\dot{\alpha}}_3}+
\frac{\tilde{\lambda}^{\dot{\alpha}}_s}{\langle n+1s \rangle}\frac{\partial}{\partial\tilde{\lambda}^{\dot{\alpha}}_{n+1}}+
\frac{\eta_{\Lambda,s}}{\langle s3 \rangle}\frac{\partial}{\partial \eta_{\Lambda,3}}+
\frac{\eta_{\Lambda,s}}{\langle n+1s \rangle}\frac{\partial}{\partial \eta_{\Lambda,n+1}}.\nonumber\\
\end{eqnarray}
Combining both terms and using Schouten identity we get
\begin{eqnarray}
\hat{S}_1^{(A_1)}+\hat{S}_1^{(A_2)}=\frac{1}{\langle sn+1\rangle}\left( \frac{\langle 1n+1\rangle}{\langle s1\rangle}
+\frac{\langle 3n+1\rangle}{\langle 3s\rangle} \right)=
\frac{\langle 13\rangle}{\langle 1s\rangle\langle s3\rangle},
\end{eqnarray}
and
\begin{eqnarray}
\hat{S}_2^{(A_1)}+\hat{S}_2^{(A_2)}=\frac{\tilde{\lambda}^{\dot{\alpha}}_s}{\langle s1 \rangle}\frac{\partial}{\partial\tilde{\lambda}^{\dot{\alpha}}_1}+
\frac{\tilde{\lambda}^{\dot{\alpha}}_s}{\langle 3s \rangle}\frac{\partial}{\partial\tilde{\lambda}^{\dot{\alpha}}_{3}}+
\frac{\eta_{\Lambda,s}}{\langle s1 \rangle}\frac{\partial}{\partial \eta_{\Lambda,1}}+
\frac{\eta_{\Lambda,s}}{\langle 3s \rangle}\frac{\partial}{\partial \eta_{\Lambda,3}}.
\end{eqnarray}
So, we see, that for the form factors of operators from $p=3$, 1/2-BPS supermultiplet soft theorem takes the form
\begin{eqnarray}
&&Z_{\bold{3},n+1}\left({\{\lambda_1,\tilde{\lambda}_1,\eta_1\},\{\epsilon\lambda_s,\tilde{\lambda}_s,\eta_s\},
\{\lambda_3,\tilde{\lambda}_3,\eta_3\},...,\{\lambda_{n+1},\tilde{\lambda}_{n+1},\eta_{n+1}\}};q,\gamma\right)=
\nonumber\\
&&\left(\frac{\hat{S}_1}{\epsilon^2}+\frac{\hat{S}_2}{\epsilon}\right)
Z_{\bold{3},n}\left({\{\lambda_1,\tilde{\lambda}_1,\eta_1\},\{\lambda_3,\tilde{\lambda}_3,\eta_3\},...,\{\lambda_{n+1},\tilde{\lambda}_{n+1},\eta_{n+1}\}};q,\gamma\right)
+reg.,~\epsilon \rightarrow0\nonumber\\
\end{eqnarray}
where
\begin{eqnarray}
\hat{S}_1=\frac{\langle 13\rangle}{\langle 1s\rangle\langle s3\rangle },
~\hat{S}_2=\frac{\tilde{\lambda}^{\dot{\alpha}}_s}{\langle s1 \rangle}\frac{\partial}{\partial\tilde{\lambda}^{\dot{\alpha}}_1}+
\frac{\tilde{\lambda}^{\dot{\alpha}}_s}{\langle s3 \rangle}\frac{\partial}{\partial\tilde{\lambda}^{\dot{\alpha}}_3}+
\frac{\eta_{\Lambda,s}}{\langle s1 \rangle}\frac{\partial}{\partial \eta_{\Lambda,1}}+
\frac{\eta_{\Lambda,s}}{\langle s3 \rangle}\frac{\partial}{\partial \eta_{\Lambda,3}}.
\end{eqnarray}

In the case of general $p$ the analysis is more involved since one have to consider residues at $ z \rightarrow \infty$ in BCFW recursion. However, using explicit answers in the MHV sector \cite{General1/2BPSformFactors}
one can verify, that universal factorization behavior holds. Thus, it is very likely that soft theorems will hold for general 1/2-BPS form factors.

Another interesting case is the form factors of operators from Konishi supermultiplet. At a moment there is no BCFW recursion available in this case and explicit answers for the form factors are known only for a limited number of external particles $n=2,3$ in MHV sector. However, as we will demonstrate in the next
section on a particular examples, universal factorization behavior holds in this case also.

The limit when (super) momentum carried by operator goes to zero $q,\gamma_+\rightarrow0$ is more involved
in this case compared to the form factors of operators from stress tensor supermultiplet. At the same time, in the case of
form factors of operators from 1/2-BPS supermultiplets $\mathcal{T}_p$ we expect that this limit is well defined.

\subsection{Some tree level examples}

Now, let us consider several explicit examples of the universal soft behavior for the form factors introduced in previous chapter. First, we will consider component version of four-point NMHV form factor of the lowest component from stress tensor operator supermultiplet $Z_{\bold{2},4}(\phi_1,\phi_2,g^-_3,g_s^+)$.

$Z_{\bold{2},4}(\phi_1,\phi_2,g^-_3,g_s^+)$ in the limit $p_s\rightarrow0$ is expected to be reduced to the $Z_{\bold{2},3}(\phi_1,\phi_2,g^-_3)$, which is NMHV 3 point form factor. An explicit expression for $Z_{\bold{2},4}(\phi_1,\phi_2,g^-_3,g_s^+)$ form factor is easily obtained using BCFW recursion
\begin{eqnarray}
Z_{\bold{2},4}(\phi_1,\phi_2,g^-_3,g_4^+)=\frac{\langle13\rangle^2}{\langle34\rangle\langle41\rangle}
\frac{1}{p_{341}^2}\frac{\langle3|p_{1234}|2]}{\langle1|p_{1234}|2]}+
\frac{[24]^2}{[23][34]}
\frac{1}{p_{234}^2}\frac{\langle1|p_{1234}|4]}{\langle1|p_{1234}|2]},
\end{eqnarray}
while for $Z_{\bold{2},3}(\phi_1,\phi_2,g^-_3)$ we have
\begin{eqnarray}
Z_{\bold{2},3}(\phi_1,\phi_2,g^-_3)=\frac{[12]^2}{[12][23][31]}.
\end{eqnarray}
It is easy to see, that
\begin{eqnarray}
\hat{S}_1Z_{\bold{2},3}(\phi_1,\phi_2,g^-_3)&=&\frac{\langle13\rangle}{\langle3s\rangle\langle s1\rangle}
\frac{[12]^2}{[12][23][31]},
\end{eqnarray}
and
\begin{eqnarray}
\hat{S}_2Z_{\bold{2},3}(\phi_1,\phi_2,g^-_3)&=&\frac{1}{\langle s1\rangle[23]}\left(\frac{[s2]}{[31]}+\frac{[12][s3]}{[31]^2}\right)
+\frac{[12]}{\langle s3\rangle}\left(\frac{[s2]}{[31][23]}
+\frac{[s1]}{[23][13]^2}\right)\nonumber\\
&=&\frac{[1s]}{\langle1s\rangle}\frac{[12]^2}{[12]^2[13]^2}+
\frac{[3s]}{\langle3s\rangle}\frac{[12]^2}{[13]^2[23]^2}.
\end{eqnarray}
Performing rescaling $\lambda_4=\lambda_s \rightarrow \epsilon \lambda_s$ for $Z_{\bold{2},4}$
\begin{eqnarray}
Z_{\bold{2},4}(\phi_1,\phi_2,g^-_3,g_s^+)=\frac{1}{\epsilon^2}\frac{\langle13\rangle^2}{\langle3s\rangle\langle s1\rangle}
\frac{1}{(p_{13}+\epsilon p_s)^2}\frac{\langle3|p_1+\epsilon p_s|2]}{\langle1|p_3+\epsilon p_s|2]}+
reg.
\end{eqnarray}
and expanding in $\epsilon$ we get
\begin{eqnarray}
Z_{\bold{2},4}(\phi_1,\phi_2,g^-_3,g_s^+) &=& \frac{1}{\epsilon^2}\frac{\langle13\rangle}{\langle3s\rangle\langle s1\rangle}
\frac{[12]^2}{[12][23][31]}+\frac{1}{\epsilon}\left(\frac{[1s]}{\langle1s\rangle}\frac{[12]^2}{[12]^2[13]^2}+
\frac{[3s]}{\langle3s\rangle}\frac{[12]^2}{[13]^2[23]^2}\right)+reg.,\nonumber\\
&& \left(\frac{\hat{S}_1}{\epsilon^2} + \frac{\hat{S_2}}{\epsilon}\right)Z_{\bold{2},3}(\phi_1, \phi_2, g_3^-) + reg. \nonumber
\end{eqnarray}
in perfect agreement with our previous considerations. Let's turn now to the form factors of operator $\mathcal{K}$ from Konishi supermultiplet:
\begin{equation}
Z_{\mathcal{K},2}(\{\lambda_1,\tilde{\lambda}_1\,\eta_1\},
\{\lambda_2,\tilde{\lambda}_2,\eta_2\};q)=
\varepsilon^{ABCD}(\eta_{A,1}\eta_{B,1})(\eta_{C,2}\eta_{D,2}),
\end{equation}
and
\begin{eqnarray}
&&Z_{\mathcal{K},3}(\{\lambda_1,\tilde{\lambda}_1\,\eta_1\},
\{\lambda_2,\tilde{\lambda}_2,\eta_2\},
\{\epsilon\lambda_s,\tilde{\lambda}_s,\eta_s\};q)=
\frac{\left(1+\mathbb{P}+\mathbb{P}^2\right)}{\langle12\rangle\langle23\rangle\langle 31\rangle}\times\nonumber\\
&\times&\left(\langle12\rangle^2\varepsilon^{ABCD}(\eta_{A,1}\eta_{B,1})(\eta_{C,2}\eta_{D,2})
+2\langle13\rangle\langle23\rangle
\varepsilon^{ABCD}(\eta_{A,1}\eta_{B,2})(\eta_{C,3}\eta_{D,3})\right).
\end{eqnarray}
As in the previous example, rescaling $\lambda_s \rightarrow \epsilon \lambda_s$ in $Z_{\mathcal{K},3}$
($[\varepsilon X] \equiv \varepsilon^{ABCD} X_{ABCD}$)
\begin{eqnarray}
&&Z_{\mathcal{K},3}
(\{\lambda_1,\tilde{\lambda}_1\,\eta_1\},
\{\lambda_2,\tilde{\lambda}_2,\eta_2\},
\{\epsilon\lambda_s,\tilde{\lambda}_s,\eta_s\};q)=\frac{1}{\epsilon^2}\frac{1}{\langle12\rangle\langle2s\rangle\langle s1\rangle}\times
\nonumber\\ &\times&\left(\langle12\rangle^2[\varepsilon(\eta_1\eta_1)(\eta_2\eta_2)]+
2\epsilon\langle12\rangle\langle1s\rangle[\varepsilon(\eta_2\eta_3)(\eta_1\eta_1)]
+2\epsilon\langle2s\rangle\langle21\rangle[\varepsilon(\eta_3\eta_1)(\eta_2\eta_2)]\right)+reg.\nonumber\\
\end{eqnarray}
and expanding in $\epsilon$ we get
\begin{eqnarray}
&&Z_{\mathcal{K},3}
(\{\lambda_1,\tilde{\lambda}_1\,\eta_1\},
\{\lambda_2,\tilde{\lambda}_2,\eta_2\},
\{\epsilon\lambda_s,\tilde{\lambda}_s,\eta_s\};q)
\nonumber\\
&=&\left(\frac{\hat{S}_1}{\epsilon^2}+\frac{\hat{S}_2}{\epsilon}\right)
Z_{\mathcal{K},2}(\{\lambda_1,\tilde{\lambda}_1\,\eta_1\},
\{\lambda_2,\tilde{\lambda}_2,\eta_2\};q)+reg.,\nonumber\\
\end{eqnarray}
where
\begin{eqnarray}
\hat{S}_1 Z_{\mathcal{K},2}
(\{\lambda_1,\tilde{\lambda}_1\,\eta_1\},
\{\lambda_2,\tilde{\lambda}_2,\eta_2\};q) &=& \frac{\langle 12\rangle}{\langle 2s\rangle\langle s1\rangle}[\varepsilon (\eta_1\eta_1)(\eta_2\eta_2)],
\nonumber \\
\hat{S}_2 Z_{\mathcal{K},2}
(\{\lambda_1,\tilde{\lambda}_1\,\eta_1\},
\{\lambda_2,\tilde{\lambda}_2,\eta_2\};q) &=& \frac{2[\varepsilon(\eta_3\eta_1)(\eta_2\eta_2)]}{\langle s1 \rangle}+\frac{2[\varepsilon(\eta_2\eta_3)(\eta_1\eta_1)]}{\langle s2\rangle}
\end{eqnarray}
and $\hat{S}_1$, $\hat{S}_2$ are given by
\begin{eqnarray}\label{S_Konishi}
\hat{S}_1 &=& \frac{\langle 12\rangle}{\langle 2s\rangle\langle s1\rangle}, \nonumber \\
\hat{S}_2 &=& \frac{\tilde{\lambda}^{\dot{\alpha}}_s}{\langle s1 \rangle}\frac{\partial}{\partial\tilde{\lambda}^{\dot{\alpha}}_1}+
\frac{\tilde{\lambda}^{\dot{\alpha}}_s}{\langle sn \rangle}\frac{\partial}{\partial\tilde{\lambda}^{\dot{\alpha}}_n}+
\frac{\eta_{A,s}}{\langle s1 \rangle}\frac{\partial}{\partial \eta_{A,1}}+
\frac{\eta_{A,s}}{\langle sn \rangle}\frac{\partial}{\partial \eta_{A,n}},
\end{eqnarray}
Here $A$ is $SU(4)_R$ index.

\subsection{Loop corrections}

At loop level, as we already mentioned in Introduction, the operators $\hat{S}_i$ may or may not receive corrections depending on the order in which soft limit and the removal of UV/IR regulator are taken. In this section we are going to consider the universal corrections to soft theorems in the case  when the removal of UV/IR regulator is taken first \cite{Bern_Soft_YM_Grav_loops}. The other case is trivial, i.e. soft theorems remain unrenormalized. So, in what follows, we consider one-loop corrections ($l=1$)
\begin{eqnarray}
\hat{S}_i=\sum_{l=0}\hat{S}_i^{(l)},~i=1,2,
\end{eqnarray}
\begin{eqnarray}
\hat{S}_1^{(0)}=\frac{\langle 1n\rangle}{\langle 1s\rangle\langle sn\rangle },
~\hat{S}_2^{(0)}=\frac{\tilde{\lambda}^{\dot{\alpha}}_s}{\langle s1 \rangle}\frac{\partial}{\partial\tilde{\lambda}^{\dot{\alpha}}_1}+
\frac{\tilde{\lambda}^{\dot{\alpha}}_s}{\langle sn \rangle}\frac{\partial}{\partial\tilde{\lambda}^{\dot{\alpha}}_n}+
\frac{\eta_{\Lambda,s}}{\langle s1 \rangle}\frac{\partial}{\partial \eta_{\Lambda,1}}+
\frac{\eta_{\Lambda,s}}{\langle sn \rangle}\frac{\partial}{\partial \eta_{\Lambda,n}}.
\end{eqnarray}
to the tree-level operators $\hat{S}_i^{(0)}$ on a few examples.

First, let us consider soft limit for the form factors of lowest component of  stress-tensor operator supermultiplet ($p=2$) in  MHV sector. These form factors are UV finite (operators from this supermultiplet are protected), but IR divergent. To regulate  IR divergences  we use dimensional regularization with $D=4-2\varepsilon$.
At tree level we have
\begin{eqnarray}
Z_{\bold{2},n}^{(0),MHV}(\{\lambda_1,\tilde{\lambda}_1\,\eta_1\},...,
\{\lambda_n,\tilde{\lambda}_n,\eta_n\};\gamma,q) &=&
\frac{\delta^{-4}(q+\gamma)\delta^{+4}(q)}{\langle12\rangle...\langle n1\rangle}.
\end{eqnarray}
Expanding Grassmann delta functions and choosing terms proportional to $(\gamma)^4$, which correspond to the projection on the lowest component of stress tensor operator supermultiplet $Tr(\phi^{++}\phi^{++})$ we get ($n=2,3$)
\begin{eqnarray}
&&Z_{\bold{2},2}^{(0),MHV}(\{\lambda_1,\tilde{\lambda}_1\,\eta_1\},
\{\lambda_2,\tilde{\lambda}_2,\eta_2\};q) = \eta_{-1,1}\eta_{-1,2}\eta_{-2,1}\eta_{-2,2},\\
&&Z_{\bold{2},3}^{(0),MHV}(\{\lambda_1,\tilde{\lambda}_1\,\eta_1\},
...,
\{\lambda_3,\tilde{\lambda}_3,\eta_3\};q) = \frac{1}{\langle 1 2 \rangle\langle 2 3\rangle\langle 31 \rangle}
 \times\nonumber \\
&\times&\Big(\langle 1 2\rangle\eta_{-1,1}\eta_{-1,2} +\langle 2 3 \rangle \eta_{-1,2}\eta_{-1,3}+\langle 1 3\rangle\eta_{-1,1}\eta_{-1,3}\Big )\times
\nonumber\\
&\times&\Big(\langle 1 2\rangle\eta_{-2,1}\eta_{-2,2} +\langle 2 3 \rangle \eta_{-2,2}\eta_{-2,3}+\langle 1 3\rangle\eta_{-2,1}\eta_{-2,3}\Big).\nonumber\\
\end{eqnarray}
One loop corrections to the above form factors are given by \cite{vanNeerven_InfraredBehaviorFormFactorsN4SYM,Brandhuber_FormFactorsN4SYM,Bork_FormFactorsN4SYM,Bork_NMHVGeneralizedUnitarity,Nandan_CuttingThroughFormFactorsN4SYM}:
\begin{eqnarray}
Z_{\bold{2},n}^{(1),MHV} &=& Z_{\bold{2},n}^{(0),MHV} f_{n}^{(1)},
\end{eqnarray}
where
\begin{eqnarray}
f_{2}^{(1)} &=& -\frac{2 c_\Gamma}{\ep^2}\left(-\frac{q^2}{\mu^2} \right)^{-\ep}, \\
f_{3}^{(1)} &=& -\frac{c_\Gamma}{\ep^2}\Big[
\left(\frac{\mu^2}{-s_{12}}\right)^{\ep}
+\left(\frac{\mu^2}{-s_{23}}\right)^{\ep}
+\left(\frac{\mu^2}{-s_{31}}\right)^{\ep} \Big. \nonumber \\
&& \Big. + \mbox{FB}(p_1,p_2,p_3,-q)  + \mbox{FB}(p_2,p_3,p_1,-q) + \mbox{FB}(p_3,p_1,p_2,-q)
\Big] ,
\end{eqnarray}
with
\begin{eqnarray}
\mbox{FB}(p_1,p_2,p_3,-q) &=& -\frac{c_\Gamma}{\ep^2}
\left[
\left(\frac{\mu^2}{-s_{12}}\right)^{\ep}h\left(-\frac{s_{31}}{s_{23}}\right)
\right.\nonumber \\ &&\left. + \left(\frac{\mu^2}{-s_{23}}\right)^{\ep}h\left(-\frac{s_{31}}{s_{12}}\right)
- \left(\frac{\mu^2}{-q^2}\right)^{\ep}h\left(-\frac{s_{31} q^2}{s_{12}s_{23}}\right)
\right]
 \\
&=&  \Li_2\left(1-\frac{q^2}{s_{12}}\right)
+ \Li_2\left(1-\frac{q^2}{s_{23}}\right) \
+ \frac{1}{2}\log^2\left(\frac{s_{12}}{s_{23}}\right)
+ \frac{\pi^2}{6} + O(\ep). \nonumber
\end{eqnarray}
Here $h(x) =~_2F_1(1,-\ep,1-\ep,x)-1$, $q^2=s_{12}+s_{23}+s_{31}$ and
\begin{eqnarray}
c_\Gamma &=& \frac{e^{\gamma_E\ep}\Gamma (1-\ep)^2\Gamma (1+\ep)}{\Gamma (1-2\ep)}
\end{eqnarray}

It is easy to verify, that at tree level in the soft limit $(\lambda_3\to \epsilon\lambda_3, \tilde{\lambda}_3\to \tilde{\lambda}_3, \eta_3\to \eta_3$) we have
\begin{eqnarray}
&& Z_{\bold{2},3}^{(0),MHV}\approx
\left(
\frac{1}{\epsilon^2}\hat{S}^{(0)}_1 + \frac{1}{\epsilon}\hat{S}^{(0)}_2
\right) Z_{\bold{2},2}^{(0),MHV} = \nonumber \\
&=& \frac{1}{\epsilon^2}\frac{\langle 12\rangle}{\langle 23\rangle\langle 31\rangle}\eta_{-1,1}\eta_{-1,2}\eta_{-1,1}\eta_{-1,2} +
\frac{1}{\epsilon}\frac{1}{\langle 31\rangle}
\Big[\eta_{-1,1}\eta_{-1,2}\eta_{-2,2}\eta_{-2,3} + \eta_{-1,2}\eta_{-1,3}\eta_{-2,1}\eta_{-2,2}\Big]
\nonumber\\&+&\frac{1}{\epsilon}\frac{1}{\langle 32\rangle}
\Big[\eta_{-1,1}\eta_{-1,2}\eta_{-2,1}\eta_{-2,3} + \eta_{-1,1}\eta_{-1,3}\eta_{-2,1}\eta_{-2,2}\Big].
\nonumber \\
\end{eqnarray}
At one loop the soft limit for the above form factor, keeping only $\log\epsilon$ enhanced terms, gives:
\begin{eqnarray}
Z_{\bold{2},3}^{(1),MHV} &\approx&
\left(\frac{1}{\epsilon^2}\hat{S}_1^{(0)} + \frac{1}{\epsilon}\hat{S}_2^{(0)} \right)F_{\bold{2},2}^{(1),MHV} + \left(\frac{1}{\epsilon^2}\hat{S}_1^{(1)} + \frac{1}{\epsilon}\hat{S}_2^{(1)} \right)F_{\bold{2},2}^{(0),MHV},
\end{eqnarray}
where
\begin{eqnarray}
\hat{S}_1^{(1)} &=& 2\log\epsilon
\left\{\frac{1}{\ep} + \log\left(-\frac{\mu^2s_{12}}{s_{13}s_{23}} \right) -\log\epsilon \right\} \hat{S}_1^{(0)}, \\
\hat{S}_2^{(1)} &=& 2\log\epsilon
\left\{\frac{1}{\ep} + \log\left(-\frac{\mu^2s_{12}}{s_{13}s_{23}} \right) -\log\epsilon \right\} \hat{S}_2^{(0)}
+ 2\log\epsilon \Big (\frac{s_{13}+s_{23}}{s_{12}}\Big)\hat{S}_1^{(0)}.
\end{eqnarray}

Next, let us consider MHV form factor of stress-tensor multiplet
at 4-point kinematics. At tree level we have
\begin{eqnarray}
&&Z_{\bold{2},4}^{(0),MHV}(\{\lambda_1,\tilde{\lambda}_1\,\eta_1\},
...,
\{\lambda_4,\tilde{\lambda}_4,\eta_4\};q) =
\frac{1}{\la 12\ra\la 23\ra\la 34\ra\la 41\ra}\times \nonumber \\
&\times& \Big(\la 12\ra\eta_{-1,1}\eta_{-1,2} + \la 13\ra\eta_{-1,1}\eta_{-1,3}
+ \la 14\ra\eta_{-1,1}\eta_{-1,4} + \la 23\ra\eta_{-1,2}\eta_{-1,3}
+ \la 24\ra\eta_{-1,2}\eta_{-1,4} \nonumber\\
&+& \la 34\ra\eta_{-1,3}\eta_{-1,4} \Big)\times \nonumber \\
&\times& \Big(
\la 12\ra\eta_{-2,1}\eta_{-2,2} + \la 13\ra\eta_{-2,1}\eta_{-2,3}
+\la 14\ra\eta_{-2,1}\eta_{-2,4} + \la 23\ra\eta_{-2,2}\eta_{-2,3}
+\la 24\ra\eta_{-2,2}\eta_{-2,4} \nonumber\\
&+& \la 34\ra\eta_{-2,3}\eta_{-2,4}
\Big)
\end{eqnarray}
One loop corrections to the above form factors are given by \cite{vanNeerven_InfraredBehaviorFormFactorsN4SYM,Brandhuber_FormFactorsN4SYM,Bork_FormFactorsN4SYM,Bork_NMHVGeneralizedUnitarity}:
\begin{eqnarray}
Z_{\bold{2},4}^{(1),MHV} &=& Z_{\bold{2},4}^{(0),MHV} f_{4}^{(1)},
\end{eqnarray}
where
\begin{eqnarray}
f_{4}^{(1)} &=& -\frac{1}{\ep^2}\sum_{l=1}^{4}
\left(-\frac{s_{ll+1}}{\mu^2} \right)^{-\ep}+  \nonumber \\
&& + \mbox{Fin}^{2me}(1,\{-q\},2,\{3,4\})
+ \mbox{Fin}^{2me}(1,2,3,\{4,-q\}) \nonumber \\
&& + \mbox{Fin}^{2me}(1,\{2,-q\},3,4) + \mbox{Fin}^{2me}(1,\{2,3\},4,\{-q\}) \nonumber \\
&& + \mbox{Fin}^{2me}(2,\{-q\},3,\{4,1\})
+ \mbox{Fin}^{2me}(2,3,4,\{1,-q\}) \nonumber \\
&& + \mbox{Fin}^{2me}(2,\{3,-q\},4,1) + \mbox{Fin}^{2me}(3,\{-q\},4,\{1,2\}).
\end{eqnarray}
Here, $\mbox{Fin}^{2me}(a,\{P\},b,\{Q\})$ denotes the finite part of two-mass easy box with massless momenta $p_a, p_b$ and corner momenta $P$ and $Q$. Expressing the two-mass easy box as a function of the kinematic invariants $s=(P+p)^2, t=(P+q)^2$ and $P^2$, $Q^2$ with $p+q+P+Q=0$, its finite part is given by
\begin{eqnarray}
\mbox{Fin}^{2me}(s,t,P^2,Q^2) = \Li_2 (1-aP^2) + \Li_2 (1-aQ^2)- \Li_2 (1-as) - \Li_2 (1-at),
\end{eqnarray}
where
\begin{eqnarray}
a = \frac{P^2+Q^2-s-t}{P^2Q^2-st}.
\end{eqnarray}

It is easy to verify, that at tree level in the soft limit $(\lambda_4\to \epsilon\lambda_4, \tilde{\lambda}_4\to \tilde{\lambda}_4, \eta_4\to \eta_4$) we have
\begin{eqnarray}
 Z_{\bold{2},4}^{(0),MHV} &\approx&
\left(
\frac{1}{\epsilon^2}\hat{S}^{(0)}_1 + \frac{1}{\epsilon}\hat{S}^{(0)}_2
\right) Z_{\bold{2},3}^{(0),MHV} =  \frac{\la 31\ra}{\epsilon^2\la 34\ra\la 41\ra} Z_{\bold{2},3}^{(0), MHV} + \frac{1}{\epsilon\la 12\ra\la 23\ra\la 34\ra\la 41\ra}\times \nonumber \\
&\times&\Big\{
[\la 14\ra\eta_{-1,1}\eta_{-1,4} + \la 24\ra\eta_{-1,2}\eta_{-1,4} + \la 34\ra\eta_{-1,3}\eta_{-1,4}] \times
\nonumber\\
&\times&[\la 12\ra\eta_{-2,1}\eta_{-2,2} + \la 13\ra\eta_{-2,1}\eta_{-2,3} + \la 23\ra\eta_{-2,2}\eta_{-2,3}]+
\nonumber\\&+& (\eta_{-1,i} \leftrightarrow \eta_{-2,i}) \Big\}.\nonumber \\
\end{eqnarray}
At one loop the soft limit for the above form factor, keeping only $\log\epsilon$ enhanced terms, gives:
\begin{eqnarray}
Z_{\bold{2},4}^{(1),MHV} &\approx&
\left(\frac{1}{\epsilon^2}\hat{S}_1^{(0)} + \frac{1}{\epsilon}\hat{S}_2^{(0)} \right)Z_{\bold{2},3}^{(1),MHV}
 + \left(\frac{1}{\epsilon^2}\hat{S}_1^{(1)} + \frac{1}{\epsilon}\hat{S}_2^{(1)} \right)Z_{\bold{2},3}^{(0),MHV},
\end{eqnarray}
where
\begin{eqnarray}
\hat{S}_1^{(1)} &=& 2\log\epsilon
\left\{\frac{1}{\ep} + \log\left(-\frac{\mu^2s_{13}}{s_{14}s_{34}} \right) -\log\epsilon \right\} \hat{S}_1^{(0)}, \\
\hat{S}_2^{(1)} &=& 2\log\epsilon
\left\{\frac{1}{\ep} + \log\left(-\frac{\mu^2s_{13}}{s_{14}s_{34}} \right) -\log\epsilon \right\} \hat{S}_2^{(0)} \nonumber \\
&& + \log\epsilon\left\{\frac{s_{14}}{s_{13}}
+ \frac{s_{24}}{s_{23}} - \frac{s_{14}s_{23}}{s_{12}s_{13}}
+ (1\leftrightarrow 3)\right\}\hat{S}_1^{(0)}.
\end{eqnarray}

It is instructive to compare, that in the case of amplitudes \cite{Bern_Soft_YM_Grav_loops}  we recover the same universal (keeping only $\log\epsilon$ enhanced terms) factor. For example, taking the soft limit $\lambda_n\to \epsilon\lambda_n, \tilde{\lambda}_n\to \tilde{\lambda}_n, \eta_n\to \eta_n$ in the case of $n$-point amplitude we have \cite{Bern_Soft_YM_Grav_loops}:
\begin{eqnarray}
\hat{S}_1^{(1)} &=& 2\log\epsilon
\left\{\frac{1}{\ep} + \log\left(-\frac{\mu^2s_{n-1,1}}{s_{n-1,n}s_{n,1}} \right) -\log\epsilon \right\} \hat{S}_1^{(0)},
\end{eqnarray}

Now let us proceed with the 3-point non-BPS form factor of Konishi supermultiplet. In contrast to 1/2-BPS form factors form factors of operators from Konishi supermultiplet
are both IR and UV divergent. Using the results from the previous subsection tree level form factors of operator $\mathcal{K}$ for particular components are given by
\begin{eqnarray}
Z_{\mathcal{K}}^{(0)}(1_{\phi_{12}}, 2_{\phi_{34}}) &=& 1, \\
Z_{\mathcal{K}}^{(0)}(1_{\phi_{12}}, 2_{\phi_{34}}, 3_{g^+}) &=&
- \frac{\langle 12\rangle}{\langle 23 \rangle\langle 31\rangle}, \\
Z_{\mathcal{K}}^{(0)}(1_{\phi_{12}}, 2_{\psi_3}, 3_{\psi_4}) &=&
\frac{1}{\langle 23\rangle}.
\end{eqnarray}
One loop corrections to the above form factors are given by \cite{Nandan_CuttingThroughFormFactorsN4SYM}:
\begin{eqnarray}
Z_{\mathcal{K}, n}^{(1)} = Z_{\mathcal{K},n}^{(0)}f_{\mathcal{K},n}^{(1)},
\end{eqnarray}
where $f_{\mathcal{K},n}^{(1)} = f_{n}^{(1)} + \tilde{f}_{\mathcal{K},n}^{(1)}$ and
\begin{eqnarray}
\tilde{f}_{K,(\phi,\phi)}^{(1)} &=& -\frac{6 c_\Gamma}{\ep (1-2\ep)}
\left(-\frac{q^2}{\mu^2}\right)^{-\ep} \\
\tilde{f}_{K,(\phi,\phi, g)}^{(1)} &=& -\frac{c_\Gamma (3+\ep)}{\ep (1-2\ep)}
\Bigg\{
\left[1+\frac{s_{13}^2+s_{23}^2}{(s_{13}+s_{23})^2} \right]
\left(\frac{\mu^2}{-q^2}\right)^{\ep}
+ \frac{2 s_{13}s_{23}}{(s_{13}+s_{23})^2}\left(\frac{\mu^2}{-s_{12}}\right)^{\ep}
\nonumber \\
&& + \frac{2\ep}{1-\ep}\frac{s_{13}s_{23}}{s_{12}(s_{13}+s_{23})^2}
\left[s_{12}\left(\frac{\mu^2}{-s_{12}}\right)^{\ep}
- q^2\left(\frac{\mu^2}{-q^2}\right)^{\ep}
\right] \Bigg\}, \\
\tilde{f}_{K, (\phi,\psi, \psi)}^{(1)} &=&
- \frac{c_\Gamma (3+\ep)}{\ep (1-2\ep)} \Bigg\{
\left[1+\frac{s_{23}-s_{12}}{s_{23}+s_{12}} \right]
\left(\frac{\mu^2}{-s_{13}}\right)^{\ep}
+ \left[1+\frac{s_{23}-s_{13}}{s_{23}+s_{13}}\right]
\left(\frac{\mu^2}{-s_{12}}\right)^{\ep} \nonumber \\
&& - \left[\frac{s_{23}-s_{12}}{s_{23}+s_{12}} + \frac{s_{23}-s_{13}}{s_{23}+s_{13}} \right]
\left(\frac{\mu^2}{-q^2}\right)^{\ep} \Bigg\}
+ \frac{3+\ep}{2}\mbox{FB}(p_3,p_1,p_2,-q).
\end{eqnarray}

Again, it is easy to see, that at tree level in the soft limit $(\lambda_3\to\epsilon\lambda_3 , \tilde{\lambda}_3\to\tilde{\lambda}_3, \eta_3\to \eta_3)$ we have
\begin{eqnarray}
&& Z_{\mathcal{K},3}^{(0)} \approx \left(\frac{1}{\epsilon^2}\hat{S}^{(0)}_1
+ \frac{1}{\epsilon}\hat{S}^{(0)}_2\right) Z_{\mathcal{K},2}^{(0)} = \nonumber \\
& -& \frac{1}{4}\sum_{A,B,C,D}\ep^{ABCD}\Big[
\frac{\langle 12\rangle}{\epsilon^2\langle 23\rangle\langle 31\rangle}
(\eta_{A,1}\eta_{B,1})(\eta_{C,2}\eta_{D,2}) -
\frac{2}{\epsilon\langle 32\rangle}(\eta_{A,1}\eta_{B,1})\eta_{C,2}\eta_{D,3}
\nonumber\\&-& \frac{2}{\epsilon\langle 31\rangle}\eta_{A,1} (\eta_{B,2}\eta_{C,2})\eta_{D,3} \Big],
\end{eqnarray}
where $\hat{S}^{(0)}_i = \hat{S}_i$ for $i=1,2$ are given by (\ref{S_Konishi}). At one loop the soft limit for the above form factor, keeping only $\log\epsilon$ enhanced terms, gives:
\begin{eqnarray}
    Z_{\mathcal{\mathcal{K}},3}^{(1)} &\approx&
    \left(\frac{1}{\epsilon^2}\hat{S}_1^{(0)} + \frac{1}{\epsilon}\hat{S}_2^{(0)} \right)Z_{\mathcal{K},2}^{(1)}  + \left(\frac{1}{\epsilon^2}\hat{S}_1^{(1)} + \frac{1}{\epsilon}\hat{S}_2^{(1)} \right)Z_{\mathcal{K},2}^{(0)},
\end{eqnarray}
where again, as in the case of 3-point MHV form factor of stress-tensor
multiplet, we have
\begin{eqnarray}
\hat{S}_1^{(1)} &=& 2\log\epsilon
\left\{\frac{1}{\ep} + \log\left(-\frac{\mu^2s_{12}}{s_{13}s_{23}} \right) -\log\epsilon \right\} \hat{S}_1^{(0)}, \\
\hat{S}_2^{(1)} &=& 2\log\epsilon
\left\{\frac{1}{\ep} + \log\left(-\frac{\mu^2s_{12}}{s_{13}s_{23}} \right) -\log\epsilon \right\} \hat{S}_2^{(0)}
+ 2\log\epsilon \Big(\frac{s_{13}+s_{23}}{s_{12}}\Big)\hat{S}_1^{(0)}.
\end{eqnarray}
So, we see, that in the case when operators $\hat{S}_i$ receive radiative corrections, the latter have universal form independent from the form factor analyzed.

\section{On-shell diagrams, Grassmannian integral and form factors}\label{p4}
\subsection{On-shell diagrams and amplitudes in $N=4$ SYM}
\begin{figure}[t]
    \begin{center}
        \leavevmode
        \epsfxsize=7cm
        \epsffile{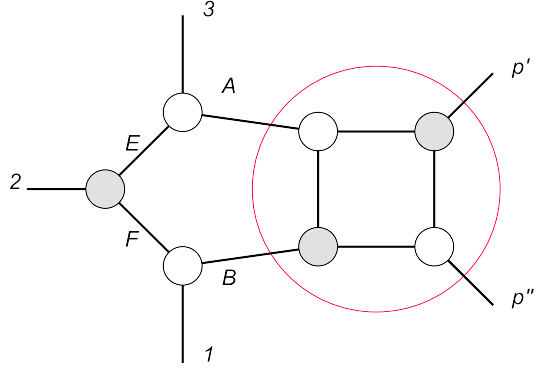}
    \end{center}\vspace{-0.2cm}
    \caption{On-shell diagram for 3-point MHV form factor (5-point MHV amplitude).
        Set of vertexes in red circle corresponds to the form factor.
        Insertion of inverse soft factor, not shown in figure,  is  assumed.}\label{MHVDeformedFormFactor}
\end{figure}
Representation of $N=4$ SYM scattering amplitudes in terms of Grassmannian integral
\cite{Arcani_Hamed_PositiveGrassmannians,Masson_Skiner_Grassmaians_Twistors,ArkaniHamed_DualitySMatrix} most
naturally incorporates all their symmetry properties, such as  Yangian invariance, and is also interesting in connection with different
twistor string theories formulations proposed recently \cite{Twistor_String_General}. Besides, it allows study of scattering amplitudes with the use of powerful tools from combinatorics and
algebraic geometry \cite{Arcani_Hamed_PositiveGrassmannians}.

Using spinor helicity variables and on-shell momentum superspace,
the Grassmannian integral representation of tree level $n$-point N$^{k-2}$MHV scattering amplitude takes the following form
\cite{ArkaniHamed_DualitySMatrix}:
\begin{eqnarray}\label{GrassmannianIntegralLambda}
    A_{n,k-2}^{(0)}(\{\lambda,\tilde{\lambda},\eta\})&=&
    \int \frac{d^{n\times k}C_{al}}{Vol[GL(k)]}\frac{1}{M_1...M_n}
    \prod_{a=1}^k
    \delta^{2}\left(\sum_{l=1}^n C_{al}\tilde{\lambda}_l\right)
    \delta^{4}\left(\sum_{l=1}^n C_{al}\eta_l\right)\times\nonumber\\
    &\times&\prod_{b=k+1}^n
    \delta^{2}\left(\sum_{l=1}^n \tilde{C}_{al}\lambda_l\right).
\end{eqnarray}
Here, $C_{al}$ is a matrix representation of $Gr(k,n)$ Grassmanian coordinates and  $M_i$ denotes $i$-th minor of $C_{al}$.. The matrix $\tilde{C}_{al}$ is given by
$$
C\tilde{C}^{T}=\sum_{i=1}^nC_{ai}\tilde{C}_{bi}=0
$$
and $k$ corresponds to the amplitude helicity configuration we are interested in, that is, $A_{3,-1}^{(0)}$ denotes $A_3^{(0)\overline{MHV}}$ amplitude,
$A_{n,0}^{(0)}$ stands for $A_3^{(0)MHV}$ amplitude and so on. In what follows,  we will also drop $^{(0)}$ superscript corresponding to the number of loops, as all objects, which we will consider in this section, will be at tree level only. The factor $Vol[GL(k)]$ in the integration measure means, that we should ``gauge fix'' arbitrary
k columns in $C_{al}$ matrix. For example, in $Gr(3,6)$
$\mbox{NMHV}$ case one can choose $GL(3)$ ``gauge'' as
\begin{eqnarray}
    C=\left( \begin{array}{cccccc}
        1 & 0 & 0 & c_{14} & c_{15} & c_{16} \\
        0 & 1 & 0 & c_{24} & c_{25} & c_{26} \\
        0 & 0 & 1 & c_{34} & c_{35} & c_{36}\end{array} \right) ,
\end{eqnarray}
so that $M_1=1$, $M_2=+c_{14}$ and so on. The integral over $d^{n\times k}C_{al}$ is understood as a multidimensional complex contour integral. The result of integration will in general depend on the choice of integration contour. Choosing different contours one can obtain different representations of the same tree level amplitude.

Now, let us discuss Grassmannian description of the amplitudes in somewhat more detail. It was shown recently, that in fact not all points of Grassmannian give nontrivial contributions to integrals above, but only those, which belong to the so called positive Grassmannian $Gr_+(k,n)$ \cite{Arcani_Hamed_PositiveGrassmannians}.
The points of Grassmannian manifold $Gr(k,n)$ are given by complex $k$-planes in $\mathbb{C}^n$ space passing through its origin. For example, the Grassmannian $Gr(1,2)$ is equivalent to complex projective space $Gr(1,2)=\mathbb{C}\mathbb{P}$. Each $k$-plane, that is a point of $Gr(k,n)$, can be parameterized by $k$  $n$-vectors in $\mathbb{C}^n$, that is by $n\times k$ matrix ($C$ matrix in Eq.
(\ref{GrassmannianIntegralLambda})). One should also take into account rotations in the $k$-plane, so that two $n\times k$ matrices, which differ only by $GL(k)$ transformation, in fact  parameterize the same point in $Gr(k,n)$. This explains $Vol[GL(k)]$ factor in the integration measure of (\ref{GrassmannianIntegralLambda}).
Positive Grassmannian $Gr(k,n)_+$ is a
submanifold in $Gr(k,n)$ defined by the condition that for every point in $Gr(k,n)_+$ with coordinates $C$ all minors $M_i$ of $C$ should be positive.

The Grassmanian $Gr(k,n)_+$ could be decomposed into the nested set of its submanifolds (called cells) depending on linear dependencies of (cyclically) consecutive column chains of $C$ (positroid stratification) \cite{Arcani_Hamed_PositiveGrassmannians}. The submanifolds (positroid cells) with larger number of linear dependent columns are being \emph{the boundaries} of submanifolds with smaller number of linear dependent columns in $C$. The submanifold
of $Gr(k,n)_+$ containing only points, whose coordinates $C$ contain no linear dependent sets of columns, is called \emph{top-cell}.

There is a correspondence between every such submanifold (\emph{positroid cell}) of $Gr(k,n)_+$ labeled by \emph{decorated permutation}\footnote{A decorated permutation is an injective map $\sigma:\{1,\ldots,n\}\mapsto\{1,\ldots,2n\}$, such that
$a\leq\sigma(a)\leq a+n$. Taking
$\sigma~\mbox{mod}~n$ will give us ordinary
permutation.} and a special diagram (\emph{on-shell diagrams}) constructed from $\mbox{MHV}_3$ (gray vertexes) and $\overline{\mbox{MHV}}_3$ (white vertexes) vertexes.
The parameters of Grassmannian $k$  and $n$ are related to the number of white $n_w$ , gray $n_g$ vertexes and number of internal lines $n_I$ as
\begin{eqnarray}
k=2n_g+n_w-n_I,~n=3(n_g+n_w)-n_I.
\end{eqnarray}
In what follows, we will also consider only on-shell diagrams (reduced graphs) with the number of faces $F$ less or equal
then the dimension of $Gr(k,n)_+$  ($dim[Gr(k,n)_+]=k(n-k)$).

Such on-shell diagrams are given by the rational functions of external kinematical data only. They are also manifestly Yangian invariant for the general set of external kinematical data \cite{Arcani_Hamed_PositiveGrassmannians}. As rational functions on-shell diagrams have poles. These \emph{poles are
in one to one correspondence with the boundaries of cells} in $Gr(k,n)_+$ to which on-shell diagrams correspond to \cite{Arcani_Hamed_PositiveGrassmannians}.
Also, there are actually equivalent classes of the on-shell diagrams which give the same rational function. Whether two on-shell diagrams are equivalent or not could be seen by considering associated permutation: equivalent diagrams have the same permutation.
There are also graphical rules (''square moves and merger''), which allow to transform one equivalent graph into another \cite{Arcani_Hamed_PositiveGrassmannians}.

The BCFW recursion for the tree-level amplitudes in this formulation could be reproduced as follows \cite{Arcani_Hamed_PositiveGrassmannians}. First, one
takes top-cell of $Gr(k,n)_+$ (the corresponding on-shell diagram could be written exactly
as in Eq. (\ref{GrassmannianIntegralLambda})) and then consider its
boundaries of certain co-dimension. The sum of the on-shell diagrams corresponding to the boundaries of the top-cell, will have only singularities corresponding to the factorization channels (poles of propagators)  of tree-level $\mbox{N}^k\mbox{MHV}$ amplitudes. The on-shell diagrams corresponding to these boundaries can be obtained from the on-shell diagram corresponding to the top-cell by removing $(k-2)(n-k-2)$ edges from it. The resulting on-shell diagrams obtained by removing edge procedure  are in one to one correspondence with the terms of ordinary BCFW recursion.

This approach, in addition to advantages mentioned in the beginning of
this section, gives a systematic procedure to prove different complicated relations between rational Yangian invariants \cite{Arcani_Hamed_PositiveGrassmannians}
(more complicated versions of "6-term identity"
\cite{Henrietta_Amplitudes,DualConfInvForAmplitudesCorch}) and helps to prove absence of spurious poles in BCFW recursion for tree level amplitudes \cite{Arcani_Hamed_PositiveGrassmannians}.
One can also use this formalism to study loop amplitudes as well ("amplituhedron"
\cite{Arcani_Hamed_PositiveGrassmannians,Arcani_Hamed_Amplituhdron_1,Arcani_Hamed_Amplituhdron_2}),
and amplitudes in theories with less supersymmetry and in different dimensions.

Now let us add a little more details concerning on-shell diagrams.
The main ingredients of the  on-shell diagrams are tree level $\mbox{MHV}_3$
and $\overline{\mbox{MHV}}_3$ amplitudes written in the form of Grassmannian integral.
For $\mbox{MHV}_3$ amplitude one gets  integral over Grassmannian $Gr(2,3)$ (see Eq. (\ref{GrassmannianIntegralLambda})):
\begin{eqnarray}
    A_{3,0}(\{\lambda,\tilde{\lambda},\eta\})&=&\int \frac{d\alpha_{1}}{\alpha_{1}}\frac{d\alpha_{2}}{\alpha_{2}}~
    \delta^{2}\left(\tilde{\lambda}_1+\alpha_1\tilde{\lambda}_3\right)
    \delta^{2}\left(\tilde{\lambda}_2+\alpha_2\tilde{\lambda}_3\right)
    \times\delta^{2}\left(\lambda_3+\alpha_1\lambda_1+\alpha_2\lambda_2\right)\times\nonumber\\
    &\times&\hat{\delta}^{4}\left(\eta_1+\alpha_1\eta_3\right)
    \hat{\delta}^{4}\left(\eta_2+\alpha_2\eta_3\right),
\end{eqnarray}
while for $\overline{\mbox{MHV}}_3$ one gets integral over Grassmannian $Gr(1,3)$
\begin{eqnarray}
    A_{3,-1}(\{\lambda,\tilde{\lambda},\eta\})&=&\int \frac{d\beta_{1}}{\beta_{1}}\frac{d\beta_{2}}{\beta_{2}}~
    \delta^{2}\left(\lambda_1+\beta_1\lambda_3\right)
    \delta^{2}\left(\lambda_2+\beta_2\lambda_3\right)
    \times\delta^{2}\left(\tilde{\lambda}_3+\beta_1\tilde{\lambda}_1+\beta_2\tilde{\lambda}_2\right)\times\nonumber\\
    &\times&
    \hat{\delta}^{4}\left(\eta_3+\beta_1\eta_1+\beta_2\eta_2\right).
\end{eqnarray}
Next, one can combine such vertexes into bigger combinations
(on-shell diagrams) connecting them by "on-shell propagators"
$$
\int\frac{d^2\lambda_I~d^2\tilde{\lambda}_I~d^4\eta_I}{U(1)}.
$$
These integrations can always be removed by corresponding delta functions and in the result one will always be left with integrals over $d\alpha/\alpha$ parameters only for any combination of on-shell vertexes. The resulting integral over $d\alpha/\alpha$ parameters for the most configurations of vertexes will have the form \cite{Arcani_Hamed_PositiveGrassmannians}
\begin{eqnarray}
    \Omega&=&\int \prod_{i=1}^{n_w}\frac{d\alpha_{1i}}{\alpha_{1i}}\frac{d\alpha_{2i}}{\alpha_{2i}}
    \prod_{j=1}^{n_g}\frac{d\beta_{1i}}{\beta_{1i}}\frac{d\beta_{2i}}{\beta_{2i}}
    \prod_{m=1}^{n_I}\frac{1}{U(1)_m}
    \times\nonumber\\
    &\times&\prod_{a=1}^k
    \delta^{2}\left(\sum_{l=1}^n C_{al}[\alpha]\tilde{\lambda}_l\right)
    \delta^{4}\left(\sum_{l=1}^n C_{al}[\alpha]\eta_l\right)\prod_{b=k+1}^n
    \delta^{2}\left(\sum_{l=1}^n \tilde{C}_{bl}[\alpha]\lambda_l\right).
\end{eqnarray}
Here, ${\alpha_{1i},\alpha_{2i},\beta_{1i},\beta_{2i}}\equiv\alpha$,
$n_w$ is the number of white vertexes, $n_g$ is the number of gray vertexes and $n_I$ is the number of internal lines in corresponding on-shell diagram. Explicit form of $C_{al}[\alpha]$ (coordinates of some cell in $Gr_+(k,n)$) in most cases  can be found by analyzing the permutation associated with the on-shell diagram \cite{Arcani_Hamed_PositiveGrassmannians}. One can think of $\Omega$ as the integral over
some differential form $d\Omega$ \cite{Arcani_Hamed_PositiveGrassmannians}. Thus, we can always think about tree (and loop) amplitudes as an integrals over
some differential form:
\begin{eqnarray}
    A_{n,k}&=&\sum_{\sigma}\int d\Omega_{\sigma}.
\end{eqnarray}
Here, $\sum_{\sigma}$ denotes the sum over the appropriate set of on-shell diagrams, which in our case are given by the  boundaries of the top-cell in $Gr(k,n)_+$.

\subsection{On-shell diagrams and form factors in $N=4$ SYM}
In this subsection we want to discuss the following observation. It turns out, that it is possible to rewrite some simple answers for form factors of stress tensor operator supermultiplet (in the case, where full answer is given by a single BCFW diagram) as some deformation of on-shell diagrams for amplitudes based on soft limit.

The idea behind this deformation is the following. As we have seen before, the amplitude is always singular in the soft limit, while the
form factor is regular in the soft limit of momentum carried by operator. Moreover, in the soft limit with respect to momentum carried by operator the form factor is very similar to the amplitude. So, the question arises - can we use this universal soft behavior to relate expressions for the form factors and amplitudes? More explicitly, we may  use inverse soft factors as regulators of the soft limit behavior with
respect to some kinematical variables $\{\lambda_i,\tilde{\lambda}_i\},\{\lambda_{i+1},\tilde{\lambda}_{i+1}\}$ in Grassmannian integral. Next, these variables could be related with the momentum $q$ carried by operator and the result of integration over
 Grassmannian with form factor.
\begin{figure}[t]
    \begin{center}
        \leavevmode
        \epsfxsize=8cm
        \epsffile{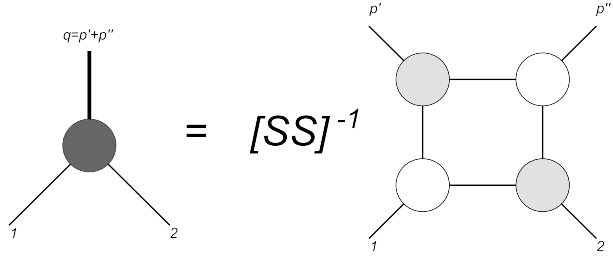}
    \end{center}\vspace{-0.2cm}
    \caption{Suggested form of the on-shell diagram vertex corresponding to form factor. The factor
$[\textbf{SS}]^{-1}$ in figure denotes the multiplication with factor $[S(2,p',p'')S(2,p'',1)]^{-1}$.}\label{FormFactorVsAmplitude}
\end{figure}
The first non-vanishing tree-level form factor is $Z_{\bold{2},2}^{MHV}$. It is easy to see, that it could be rewritten as\footnote{Here, for saving space we use abbreviation $\{\lambda_i,\tilde{\lambda}_i,\eta_i\}\equiv i$.}
\begin{eqnarray}
    Z_{\bold{2},2}^{MHV}(1,2)=[S(2,p',p'')S(2,p'',1)]^{-1}A_{4,0}(1,2,p',p'').
\end{eqnarray}
\begin{eqnarray}
    S(i,s,j)=\frac{\langle ij\rangle}{\langle is\rangle\langle sj\rangle},~q=p'+p'',~p'^2=p''^2=0.
\end{eqnarray}
Here, the form factor momentum $q$ was split into two lightlike vectors, for which spinor helicity representation was used. This decomposition is similar to those used in the applications of spinor helicity formalism to the gauge theories with spontaneously broken symmetry \cite{HenrietaConstructingAmplInGeneralTheories}. Now, we can use representation of $A_{4,0}(1,2,p',p'')$ amplitude in terms of the integral over
$Gr(2,4)$ Grassmannian and use this block in the on-shell diagrams for the form factors  (see Fig.~\ref{MHVDeformedFormFactor}).

In the case when the momentum carried by operator is lightlike ($q^2=0$)
the situation is even more simple and we have
\begin{eqnarray}
    Z_{\bold{2},2}^{MHV}(1,2)=[S(1,q,2)]^{-1}A_{3,0}(1,2,q).
\end{eqnarray}
So, here we can use essentially the same MHV vertex as in the case of amplitudes, but now with additional factor $[S(1,q,2)]^{-1}$.
\begin{figure}[t]
    \begin{center}
        \leavevmode
        \epsfxsize=7cm
        \epsffile{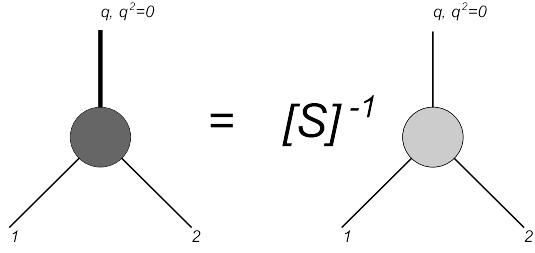}
    \end{center}\vspace{-0.2cm}
    \caption{Suggested form of the on-shell diagram vertex corresponding to form factor in the case of the lightlike momentum  $q^2=0$. The factor $[\textbf{S}]^{-1}$ in figure denotes the multiplication with factor $[S(1,q,2)]^{-1}$.}\label{FormFactorVsAmplitudeZero}
\end{figure}

Now, let us consider several simple examples of on-shell diagrams with
above mentioned representation for the form factor inserted. Here, we are
considering only cases, when full answer for form factor is given by one diagram. Also here we will not discuss combinatorial properties of form factor on-shell diagrams or their equivalence relations with respect to square move or merger transformation rules. Most likely, these transformation rules will not generate equivalent diagrams any more. The situation is similar to the one in non-planar case for the amplitudes \cite{Positive_Grassmannian_none_planar}.
Note also, that due to their color structure, form factors beyond tree-level will also contain non-planar contributions even in the planar limit. So, such behavior of the deformed on-shell diagrams may be reasonable.

First of all, consider $\mbox{MHV}_n$ form factor of stress tensor operator supermultiplet corresponding to $Gr(2,n)$ integral. The integral representation for the $\mbox{MHV}_{n+2}$ on-shell diagram with the described above deformation takes the form\footnote{Here, we omit irrelevant for our discussion Grassmann delta functions.}:
\begin{eqnarray}
    \Omega&=&\int \prod_{i=1}\frac{d\alpha_{1i}}{\alpha_{1i}}
    ...\frac{d^2\lambda_Ad^2\tilde{\lambda}_A}{U(1)}\frac{d^2\lambda_Bd^2\tilde{\lambda}_B}{U(1)}...
    \frac{\langle Ap'\rangle\langle p'p''\rangle\langle p''B\rangle}{\langle AB\rangle}...
    \times\nonumber\\&\times&\delta^{2}\left(\tilde{\lambda}_n+\beta_{A}\tilde{\lambda}_A+\beta_{E}\tilde{\lambda}_E\right)
    \delta^{2}\left(\lambda_A-\beta_A\lambda_n\right)
    \delta^{2}\left(\lambda_E-\beta_E\lambda_n\right)
    \times\nonumber\\&\times&\delta^{2}\left(\tilde{\lambda}_1+\beta_{B}\tilde{\lambda}_B+\beta_{F}\tilde{\lambda}_F\right)
    \delta^{2}\left(\lambda_B-\beta_B\lambda_1\right)
    \delta^{2}\left(\lambda_F-\beta_F\lambda_1\right)...~.
\end{eqnarray}
\begin{figure}[t]
    \begin{center}
        \leavevmode
        \epsfxsize=7cm
        \epsffile{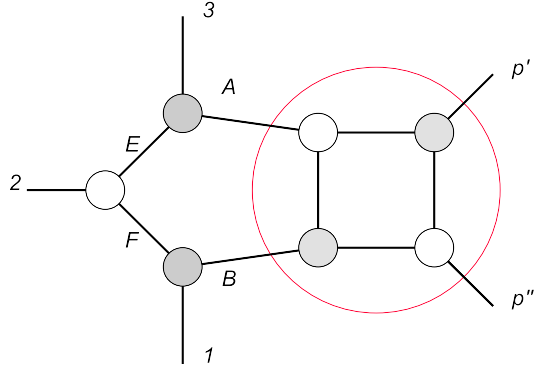}
    \end{center}\vspace{-0.2cm}
    \caption{On-shell diagram for 3-point NMHV form factor.
        Set of vertexes in red circle corresponds to the form factor.
        Insertion of inverse soft factor, not shown in figure, is assumed.}\label{NMHVDeformedFormFactor}
\end{figure}
Here, we have written  explicitly inverse soft factor contribution together with contributions from $\overline{\mbox{MHV}}_3$ vertexes $(E,n,A)$ and $(B,1,F)$ (see Fig.~\ref{MHVDeformedFormFactor} for $n=3$ example). One can see, that after integration with respect to $d^2\lambda_Ad^2\tilde{\lambda}_A$ and $d^2\lambda_Bd^2\tilde{\lambda}_B$ on the support of corresponding delta functions $\lambda_B=\beta_B\lambda_1$ and $\lambda_A=\beta_A\lambda_n$. Thus, the inverse soft factor reduces to
$$
\frac{\langle Ap'\rangle\langle p'p''\rangle\langle p''B\rangle}{\langle AB\rangle} \rightarrow\frac{\langle 1p'\rangle\langle p'p''\rangle\langle p''n\rangle}{\langle 1n\rangle},
$$
and can be moved away from integral sign. The rest of the diagram is just $\mbox{MHV}_{n+2}$
amplitude $A_{n+2,0}(1,...n,p',p'')$. As the result we get:
\begin{eqnarray}
    \Omega=\frac{\langle 1p'\rangle\langle p'p''\rangle\langle p''n\rangle}{\langle 1n\rangle}
    A_{n+2,0}(1,...n,p',p'')=Z_{\bold{2},n}^{MHV}(1,...,n),
\end{eqnarray}
as expected. The same will be true for the $q^2=0$ case as well.

Now, consider the case of $\mbox{NMHV}_3$ form factor corresponding to
$Gr(3,5)$ integral. Again, we are going to consider
on-shell diagram for $\mbox{NMHV}_5$ amplitude with deformation and expect, that such integral will
give us $\mbox{NMHV}_3$ form factor. So, we have:
\begin{eqnarray}
    \Omega&=&\int \prod_{i=1}\frac{d\alpha_{1i}}{\alpha_{1i}}
    ...\frac{d^2\lambda_Ad^2\tilde{\lambda}_A}{U(1)}\frac{d^2\lambda_Bd^2\tilde{\lambda}_B}{U(1)}...
    \frac{\langle Ap'\rangle\langle p'p''\rangle\langle p''B\rangle}{\langle AB\rangle}...
    \times\nonumber\\&\times&\delta^{2}\left(\lambda_3+\alpha_{A}\lambda_A+\alpha_{E}\lambda_E\right)
    \delta^{2}\left(\tilde{\lambda}_A-\alpha_A\tilde{\lambda}_3\right)
    \delta^{2}\left(\tilde{\lambda}_E-\alpha_E\tilde{\lambda}_3\right)
    \times\nonumber\\&\times&\delta^{2}\left(\lambda_1+\alpha_{B}\lambda_B+\alpha_{F}\lambda_F\right)
    \delta^{2}\left(\tilde{\lambda}_B-\alpha_B\tilde{\lambda}_1\right)
    \delta^{2}\left(\tilde{\lambda}_F-\alpha_F\tilde{\lambda}_1\right)\times
    \nonumber\\&\times&\delta^{2}\left(\tilde{\lambda}_2+\beta_{E}\tilde{\lambda}_E+\beta_{F}\tilde{\lambda}_F\right)
    \delta^{2}\left(\lambda_E-\beta_E\lambda_2\right)
    \delta^{2}\left(\lambda_F-\beta_F\lambda_2\right).
\end{eqnarray}
Integrating with respect to $A$ and $B$ internal lines we get
$$
\lambda_B\sim\lambda_1+c_1\lambda_2,~\lambda_A\sim\lambda_3+c_2\lambda_2,
$$
with
$c_1=\alpha_F\beta_E$, $c_2=\lambda_E\beta_F$.
We can continue taking integrals with respect to internal
lines and $\alpha$'s and $\beta$'s. At the end of the day, we get
\begin{eqnarray}
    \Omega=\frac{\langle Ap'\rangle\langle p'p''\rangle\langle p''B\rangle}{\langle AB\rangle}
    A_{5,1}(1,2,3,p',p''),
\end{eqnarray}
where $c_1=[23]/[13]$ and $c_2=[12]/[31]$. This expression could be further  transformed to the $Z_{\bold{2},3}^{NMHV}(1,2,3)$
form factor ($q^2=\langle p'p''\rangle[p'p'']$ and the notations  $q_{1...n}\equiv\sum_{i=1}^n\lambda_i\eta_i$ and $p_{1...n}\equiv\sum_{i=1}^np_i$ are used). Indeed, we have
\begin{eqnarray}
    \Omega=\frac{\langle Ap'\rangle\langle p'p''\rangle\langle p''B\rangle}{\langle AB\rangle}
    \frac{\delta^4(p_{123}+q)\delta^8(q_{123}+\gamma)\hat{\delta}^4([12]\eta_3+\mbox{perm.})}
    {\langle p'p''\rangle^4[12][23][3p'][p'p''][p''1]}.\nonumber\\
\end{eqnarray}
Dropping total momentum conservation delta function, we can write
\begin{eqnarray}
    \Omega=\frac{\langle Ap'\rangle\langle p'p''\rangle\langle p''B\rangle}{\langle AB\rangle}
    \frac{[13][p'p'']}{[3p'][p''1]\langle p'p''\rangle^2}
    \frac{\delta^8(q_{123}+\gamma)\hat{\delta}^4([12]\eta_3+\mbox{perm.})}
    {[12][23][31]q^4}.\nonumber\\
\end{eqnarray}
Substituting explicit expressions for $\lambda_B$ and $\lambda_A$,  the $\lambda_A,\lambda_B$ dependent prefactor takes the form
\begin{eqnarray}
    \frac{\langle Ap'\rangle\langle p'p''\rangle\langle p''B\rangle}{\langle AB\rangle}
    =[13]^2\langle Ap'\rangle\langle p''B\rangle\frac{\langle p'p''\rangle}{[13]^2\langle AB\rangle}=
    \frac{\langle p'|1+2|3]\langle p''|3+2|3]}{[13][p'p'']},\nonumber\\
\end{eqnarray}
which can be further transformed using momentum conservation to
\begin{eqnarray}
    \frac{\langle Ap'\rangle\langle p'p''\rangle\langle p''B\rangle}{\langle AB\rangle}
    =\frac{\langle p'p''\rangle^2[3p'][p''1]}{[13][p'p'']}.
\end{eqnarray}
Keeping in mind, that \cite{Bork_NMHVGeneralizedUnitarity}
\begin{eqnarray}
    Z^{NMHV}_{\bold{2},3}(1,2,3)=
    \frac{\delta^8(q_{123}+\gamma)\hat{\delta}^4([12]\eta_3+\mbox{perm.})}
    {[12][23][31]q^4},\nonumber\\
\end{eqnarray}
we see, that indeed the expected identity $\Omega=Z^{NMHV}_{\bold{2},3}(1,2,3)$ holds true.

So, on these simple examples we see, that suggested here
modification of the on-shell diagrams gives reasonable
results for some simple examples. It is still unclear for us whether it is possible to write some consistent deformation of Grassmannian integral (\ref{GrassmannianIntegralLambda}),
which will give us tree-level form factors
of stress tensor operator supermultiplet.
Relations between different terms of BCFW and BCFW/CSW recursion
observed in \cite{BORK_POLY} suggest that such deformation exists.
It is likely, that it would be easier to find such deformation in the case when momentum $q$ carried by operator is also lightlike $q^2=0$.

\subsection{BCFW for form factors of stress tensor supermultiplet with $q^2=0$, $\mbox{N}\mbox{MHV}$ sector}
Here, we would like to discuss BCFW relations for the form factors of stress tensor operator supermultiplet with $q^2=0$, which we will need when discussing Yangian symmetry properties of form factors in the limit $q^2\to 0$. MHV sector is identical to $q^2 \neq 0$ case. The
$n=3$ point form factor is equal to zero starting from NHMV sector for
the same reasons as 4-point NMHV amplitude. First non-trivial contributions in the NMHV sector start from $n=4$ case. For the $[1,2 \rangle$ shift we have
\begin{eqnarray}
    Z_{\bold{2},4}^{NMHV}=Z_{\bold{2},4}^{MHV}
    \left(R^{(1)}_{132}+R^{(2)}_{142}\right),
\end{eqnarray}
where $R^{(1)}_{rst}$ and $R^{(2)}_{rst}$ functions are given by \cite{Bork_NMHVGeneralizedUnitarity}
\begin{figure}[t]
    \begin{center}
        \epsfxsize=6cm
        \epsffile{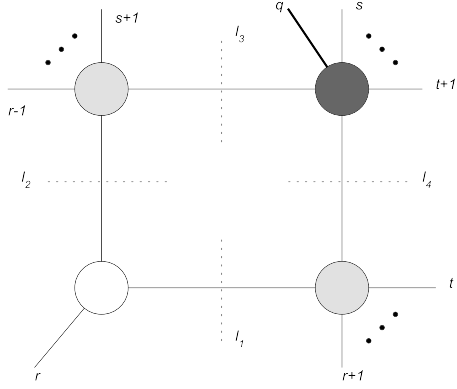}
    \end{center}\vspace{-0.2cm}
    \caption{Diagrammatic representation of the
        quadruple cut proportional to $R^{(1)}_{rst}$.
        The dark grey blob is the MHV form factor.}\label{NMHV_R1}
\end{figure}
\begin{figure}[t]
    \begin{center}
        \epsfxsize=6cm
        \epsffile{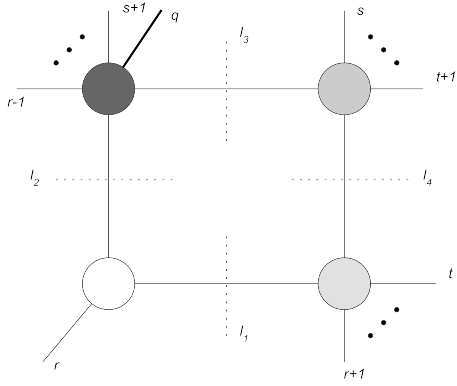}
    \end{center}\vspace{-0.2cm}
    \caption{Diagrammatic representation of the quadruple cut proportional to $R^{(2)}_{rst}$. }\label{NMHV_R2}
\end{figure}
\begin{figure}[h]
 \begin{center}
  \epsfxsize=6cm
 \epsffile{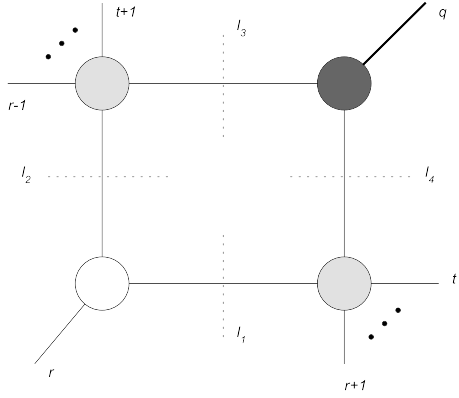}
 \end{center}\vspace{-0.2cm}
 \caption{Diagrammatic representation of the quadruple cut proportional to $\tilde{R}^{(1)}_{rtt}$. }\label{NMHV_R1lim}
 \end{figure}
\begin{eqnarray}\label{R_1_deff}
    R_{rst}^{(1)}&=&\frac{\langle s+1s\rangle\langle
        t+1t\rangle\hat{\delta}^4\left(\sum_{i=t}^{r+1}\eta_i\langle
        i|p_{t...s+1}p_{r...s+1}|r\rangle-\sum_{i=r}^{s+1}\eta_i\langle
        i|p_{t...s+1}p_{t...r+1}|r\rangle\right)} {p_{s+1...t}^2\langle
        r|p_{r...s+1}p_{t...s+1}|t+1\rangle\langle
        r|p_{r...s+1}p_{t...s+1}|t\rangle\langle
        r|p_{t...r+1}p_{t...s+1}|s+1\rangle\langle
        r|p_{t...r+1}p_{t...s+1}|s\rangle},\nonumber\\
\end{eqnarray}
\begin{eqnarray}\label{R_2_deff}
    R_{rst}^{(2)}&=&\frac{\langle s+1s\rangle\langle
        t+1t\rangle\hat{\delta}^4\left(\sum_{i=t}^{r+1}\eta_i\langle
        i|p_{s...t+1}p_{s...r+1}|r\rangle-\sum_{i=r+1}^{s}\eta_i\langle
        i|p_{s...t+1}p_{t...r+1}|r\rangle\right)} {p_{s...t+1}^2\langle
        r|p_{s...r+1}p_{s...t+1}|t+1\rangle\langle
        r|p_{s...r+1}p_{s...t+1}|t\rangle\langle
        r|p_{t...r+1}p_{s...t+1}|s+1\rangle\langle
        r|p_{t...r+1}p_{s...t+1}|s\rangle}.\nonumber\\
\end{eqnarray}
In fact, these functions coincide with $R_{rst}$ dual conformal
invariants, when rewritten with the use of momentum conservation
in such a way, that the dependence on variables associated
with upper right (left) corner of corresponding diagram (see Figs. \ref{NMHV_R1} and \ref{NMHV_R2})
is dropped. There is also special case of $R_{rst}^{(1)}$, which
we denote as $\tilde{R}_{rtt}^{(1)}$. We will need its expression in the next section and thus will also write it here for completeness:
\begin{eqnarray}\label{tildeR_1_deff}
\tilde{R}_{rtt}^{(1)}&=&\frac{\langle
tt+1\rangle\hat{\delta}^4\left(\sum_{i=t}^{r+1}\eta_i\langle
i|qp_{r...t+1}|r\rangle-\sum_{i=r}^{t+1}\eta_i\langle
i|qp_{t...r+1}|r\rangle\right)}{q^4\langle
r|p_{r...t+1}q|t\rangle\langle r|p_{t...r}q|t+1\rangle
\langle r|p_{r...t}q|r\rangle}.\nonumber\\
\end{eqnarray}
\begin{figure}[t]
    \begin{center}
        \epsfxsize=14cm
        \hbox{\hspace*{3cm}
        \epsffile{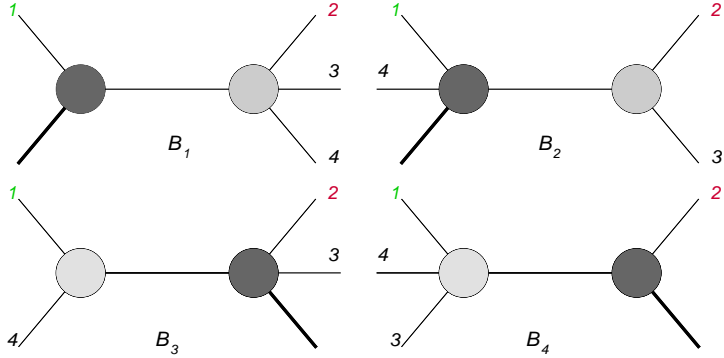}}
    \end{center}\vspace{-0.1cm}
    \caption{BCFW diagrams contributing to the $n=4$ case
        for the $[\textcolor{green}{1},\textcolor{red}{2} \rangle$ shift.
        $B_2=B_4=0$ due to the kinematical reasons.}\label{NMHV4q0}
\end{figure}
It is possible to obtain the solution of BCFW recursion in NMHV sector for general $n$ and it reads as:
\begin{eqnarray}
    Z_{\bold{2},n}^{NMHV}=Z_{\bold{2},n}^{MHV}\left(\sum_{i=2}^{n-2}\sum_{j=i+1}^{n-1}R^{(1)}_{1ji}+
    \sum_{i=2}^{n-2}\sum_{j=i+2}^{n}R^{(2)}_{1ji}\right).
\end{eqnarray}
Here, we would like to note, that  $\tilde{R}_{rtt}^{(1)}$ functions,
which were present in the $q^2 \neq 0$ case, are absent now.
Similar expression can be obtained for
$\mbox{N}^k\mbox{MHV}$ sectors as well. They have the simplest
form in the momentum twistor space. Details can be found in appendix C.

\subsection{Comments on Yangian invariance of tree-level form factors at $q^2 = 0$}
As we have seen already, the results obtained for tree-level MHV and NMHV form factors of stress tensor operator supermultiplet are very similar to those in the case of scattering amplitudes. It was also shown in previous subsection, that there may be some generalization (deformation) of on-shell diagram formalism for the case of form factors. On the other hand, on-shell diagrams are tightly related to the Grassmannian integral representation of Yangian invariants. So, it is natural to ask the question ``what are the properties of form factors with
respect to dual conformal (Yangian) symmetry transformations ?''.
Here, we want to share the following observations:

\begin{enumerate}
	\item MHV tree level form factors at $q^2=0$ transform covariantly with respect to dual conformal transformations \cite{DualConfInvForAmplitudesCorch} $K^{\alpha\dot{\alpha}}$ (see appendix \ref{aB})
	\begin{eqnarray}
	K^{\alpha\dot{\alpha}}Z_{\bold{2},n}^{MHV}(1,\ldots,n)=-(\sum_{i=1}^{n}x_i^{\alpha\dot{\alpha}})Z_{\bold{2},n}^{MHV}(1,\ldots,n),
	\end{eqnarray}
	and are likely to be annihilated by other generators of dual (super)conformal algebra\footnote{At least before the limit $\gamma_-\to 0$ is taken. This is, however, is likely unimportant and related to the chiral truncation of stress tensor operator supermultiplet considered in this paper. This feature should be absent in non-chiral formulation.}
	\item Using momentum supertwistors defined on periodical contour
	it is possible, at least formally,
	to rewrite ratios of $\mbox{N}^k\mbox{MHV}$ to $\mbox{MHV}$ form factors  with $q^2=0$ as a sum
	of products of $[a,b,c,d,e]$ dual conformal (Yangian) invariants
	\cite{Masson_Skiner_Grassmaians_Twistors} (see appendix \ref{aC}).
	For example, in the case of $n=4,5$ for NMHV sector we get:
	\begin{eqnarray}
	\frac{Z_{\bold{2},4}^{NMHV}}{Z_{\bold{2},4}^{MHV}}=[1,2,3,4,5]+[1,2,3,0,-1],
	\end{eqnarray}
	\begin{eqnarray}
	\frac{Z_{\bold{2},5}^{NMHV}}{Z_{\bold{2},5}^{MHV}}
	&=&[1,3,4,5,6]+[1,2,3,4,5]+[1,2,3,5,6]\nonumber\\
	&+&[-1,0,1,3,4]+[1,2,3,-2,-1]+[1,2,3,-1,0].
	\end{eqnarray}
\end{enumerate}

These observations suggest, that the form factors of operators
from stress tensor supermultiplet with $q^2=0$ exhibit dual (super)conformal and possibly Yangian invariance. Here we would like to note, that the dual conformal invariance of the form factors at light-cone, considered in \cite{Derkachov_DualConformalSymmetryLightcone} is the special case of the limit $q^2\to 0$ considered in our paper.
The  Yangian symmetry in its turn is usually an indication for the presence of some integrable structure. We are going to discuss this possibility in the next section.

\section{Inverse soft limit and integrability}\label{p5}
\subsection{Inverse soft limit and integrability for amplitudes}
All types of recursion relations for scattering amplitudes (BCFW, all-line shift (CSW) and so on) employ the knowledge of amplitude singularity structure together with amplitude factorization properties near such singularities. It is natural to ask whether the soft behavior of scattering amplitudes are also just enough to restrict the structure of amplitudes? The answer to this question is positive. At a moment there is a well known Inverse Soft Limit (ISL) iterative procedure proposed in \cite{ArkaniHamed_DualitySMatrix}, elaborated in \cite{Bullimore_ISF_GrassmanianResidues} and applied later in \cite{Nandan_GeneretingTreeAmplitudes_ISL,BoucherVeronneau_Amplitudes_ISL} to reconstruct BCFW terms for arbitrary tree level amplitudes starting with 3-point amplitude and inserting at each step one additional external state. Moreover, from the point of view of the Grassmannian, ISL \cite{ArkaniHamed_AllLoopIntegrand,ArkaniHamed_UnificationResidues} turns out to be a natural way of constructing Yangian-invariants \cite{DualConfInvForAmplitudesCorch,Drummond_YangianSymmetryAmplitudes}, which is expected since tree-level amplitudes in $\mathcal{N}=4$ SYM are Yangian invariant.

Any tree-level super amplitude in $\mathcal{N}=4$ SYM theory could be generated with the use of ISL procedure and the result can be schematically\footnote{We refer the reader to \cite{Nandan_GeneretingTreeAmplitudes_ISL} for details of how the external states should be added to the original 3-point amplitude to obtain BCFW amplitude of interest.} written as \cite{Nandan_GeneretingTreeAmplitudes_ISL}:

\begin{eqnarray}
A_n=\sum_{i;R,L}(\prod_{L}\mathcal{S}_L)(\prod_{R}\mathcal{S}_R)
A_3^{\overline{MHV}}(i,i+1,n)|_{subst.},
\end{eqnarray}
where soft factors $\mathcal{S}_{L,R}$ are equal to either $\mathcal{S}_+$ or $\mathcal{S}_-$ defined below and $|_{subst.}$ subscript means that one has to make several folded substitutions for
spinors $\lambda_i$, $\tilde{\lambda}_i$ and Grassmann variables $\eta_i$ which are similar to BCFW shifts in 2 particle channels. Namely, we will consider 2 types of substitutions (shifts). One of them we will call negative shift and denote by $\hat{i}^{-}$
\begin{eqnarray}
&&\hat{\lambda}_{i-1}=\lambda_{i-1}+\frac{[ i+1i]}{[i-1i+1]}\lambda_{i},\nonumber\\
&&\hat{\lambda}_{i}=\lambda_{i},\nonumber\\
&&\hat{\lambda}_{i+1}=\lambda_{i+1}+\frac{[ i-1i]}{[i-1i+1]}\lambda_{i},
\nonumber\\
&&\hat{\tilde{\lambda}}_{i-1}=\hat{\tilde{\lambda}}_{i-1},\nonumber\\
&&\hat{\tilde{\lambda}}_{i}=\hat{\tilde{\lambda}}_{i},\nonumber\\
&&\hat{\tilde{\lambda}}_{i+1}=\hat{\tilde{\lambda}}_{i+1},\nonumber\\
&&\hat{\eta}_{i-1}=\eta_{i-1},\nonumber\\
&&\hat{\eta}_{i}=\eta_{i},\nonumber\\
&&\hat{\eta}_{i+1}=\eta_{n+1},.
\end{eqnarray}
and the other one will be called positive shift and denoted by $\hat{i}^{+}$
\begin{eqnarray}
&&\hat{\lambda}_{i-1}=\lambda_{i-1},\nonumber\\
&&\hat{\lambda}_{i}=\lambda_{i},\nonumber\\
&&\hat{\lambda}_{i+1}=\lambda_{i+1},\nonumber\\
&&\hat{\tilde{\lambda}}_{i-1}=\tilde{\lambda}_{i-1}+\frac{\langle i+1i\rangle}{\langle i-1i+1\rangle}\tilde{\lambda}_{i},\nonumber\\
&&\hat{\tilde{\lambda}}_{i}=\tilde{\lambda}_{i},\nonumber\\
&&\hat{\tilde{\lambda}}_{i+1}=\tilde{\lambda}_{i+1}+\frac{\langle i-1i \rangle}{\langle i-1i+1 \rangle}\tilde{\lambda}_{i},\nonumber\\
&&\hat{\eta}_{i-1}=\eta_{i-1}+\frac{\langle i+1i\rangle}{\langle i-1i+1\rangle}\eta_{i},\nonumber\\
&&\hat{\eta}_{i}=\eta_{i},\nonumber\\
&&\hat{\eta}_{i}=\eta_{i+1}+\frac{\langle i-1i \rangle}{\langle i-1i+1 \rangle}\eta_{i},
\end{eqnarray}
So defined positive and negative shifts correspond to respectively $k$ preserving and $k$-increasing inverse soft operations, here $k$ denotes the degree of $R$-charges of N$^k$MHV amplitudes:
\begin{eqnarray}
&& \mathcal{S}_{+}(i-1ii+1)A_{n-1,k}(...\widehat{(i-1)}^{+},\widehat{(i+1)}^{+},...)
= A_{n,k}(...,i-1,i,i+1,...), \nonumber \\
&& \mathcal{S}_{-}(i-1ii+1)A_{n-1,k-1}(...\widehat{(i-1)}^{-},\widehat{(i+1)}^{-},...)
= A_{n,k}(...,i-1,i,i+1,...), \nonumber \\
\end{eqnarray}
where soft factors $\mathcal{S}_+$ and $\mathcal{S}_-$ are given by
\begin{eqnarray}
&&\mathcal{S}_+(i-1,i,i+1)=\frac{\langle i-1i+1 \rangle}{\langle i-1i \rangle\langle ii+1 \rangle},\nonumber\\
&&\mathcal{S}_-(i-1,i,i+1)=
\frac{\hat{\delta}^4(\eta_{i-1}[ii+1]+\mbox{perm}.)}{[ii-1][i+1i][i-1i+1]^3}.
\end{eqnarray}
Here we would like to note, that in practice expressions generated by these substitutions could be rather complicated.

There is also a nice connection between the ISL iterative procedure described above and Quantum Inverse Scattering Method (QISM) used to construct Yangian invariants relevant for the tree-level scattering amplitudes of $\mathcal{N}=4$ SYM \cite{Frassek_BetheAnsatzYangianInvariants,Chicherin_YangBaxterScatteringAmplitudes}. Within QISM framework it was proposed to study certain auxiliary spin chain monodromies build from local Lax operators. Yangian invariants and thus amplitudes are then found as the eigenstates of these monodromies. Further, in \cite{Kanning_ShortcutAmplitudesIntegrability,Broedel_DictionaryRoperatorsOnshellGraphsYangianAlgebras} a systematic classification of Yangian invariants obtained within QISM was provided. Yangian invariance can be defined in a very compact form as a system of eigenvalue equations for the elements of a suitable monodromy matrix $M(u)$ \cite{Frassek_BetheAnsatzYangianInvariants,Chicherin_YangBaxterScatteringAmplitudes,Kanning_ShortcutAmplitudesIntegrability}:
\begin{eqnarray}
M_{ab}(u)|\Psi\rangle = {C}_{ab}|\Psi\rangle,
\end{eqnarray}
where $u$ is the spectral parameter, $C_{ab}$ are monodromy eigenvalues and monodromy eigenvectors $|\Psi\rangle$ are elements of the Hilbert space $V = V_1\otimes\ldots\otimes V_n$ with $V_i$ being a representation space of a particular $\mathfrak{gl}(N|M)$ representation. To describe tree-level scattering amplitudes one will need to specialize to the case of $N|M = 4|4$ and its non-compact representations  build using a single family of Jordan-Schwinger harmonic superoscillators $\overline{\bf w}_a, {\bf w}_b$, $a,b = 1\ldots N+M$. The latter could be conveniently realized in terms of "supertwistor" variables $\mathcal{W}_a$:
\begin{eqnarray}
J_{ab} = \overline{\bf w}_a {\bf w}_b,\quad \overline{\bf w}_a = {\mathcal W}_a,\quad {\bf w}_a = \partial_{\mathcal{W}_a}\quad  \mbox{and} \quad [{\bf w}_a, \overline{\bf w}_b\} = \delta_{ab}
\end{eqnarray}
A vacuum state for the Hilbert space $V$ used to construct Yangian invariants $|\Psi\rangle_{n,k}$ corresponding to N$^k$MHV $n$-point tree-level amplitudes $A_{n,k}$ is given by
\begin{eqnarray}
|\bf 0\rangle = \prod_{i=1}^k\delta^{N|M}(\mathcal{W}^i).
\end{eqnarray}
The monodromy matrix of the auxiliary spin chain reads
\begin{eqnarray}
M(u) = \mathcal{L}_1 (u,v_1)\ldots\mathcal{L}_k (u,v_k)
\mathcal{L}_{k+1} (u,v_{k+1})\ldots\mathcal{L}_n (u,v_n),
\end{eqnarray}
 where $u$ is again the spectral parameter, $v_i$ are spin chain inhomogeneities and Lax operators $\mathcal{L}_i (u,v)$ are given by
 \begin{eqnarray}
 \mathcal{L} (u,v) = u - v + \sum_{a,b} e_{ab} J_{ba} \label{LaxOperator}
 \end{eqnarray}
 The Bethe ansatz solution of the above spin chain leads to the following expressions for Yangian invariants labeled by the permutations $\sigma$ with minimal\footnote{The decomposition is minimal in a sense, that there exists no other decomposition of $\sigma$ into a smaller number of transpositions.} decomposition $\sigma = (i_P,j_P)\ldots (i_1,j_1)$ \cite{Kanning_ShortcutAmplitudesIntegrability}
 \begin{eqnarray}
 |\Psi\rangle = \mathcal{R}_{i_1,j_1}(\bar{u}_1)\ldots\mathcal{R}_{i_Pj_P}(\bar u_P)|\bf 0\rangle
 \end{eqnarray}
 with \cite{Chicherin_YangBaxterScatteringAmplitudes} (see also \cite{Frassek_BetheAnsatzYangianInvariants})
 \begin{eqnarray}
 \mathcal{R}_{ij} (u) = (-\overline{\bf w}^j\cdot {\bf w}^i)^u = -\frac{\Gamma (u+1)}{2\pi i}\int_{\mathcal{C}}\frac{d\alpha}{(-\alpha)^{1+u}}e^{\alpha\overline{\bf w}^j\cdot {\bf w}^i}
 \end{eqnarray}
 and
 \begin{eqnarray}
 \bar u_p = y_{\tau_p (j_p)} - y_{\tau_p} (i_p),\quad y_{\sigma (i)} + \mathfrak{s}_{\sigma(i)}=y_i,\quad  \tau_p=\tau_{p-1}\circ(i_p,j_p)=(i_1,j_1)\cdots(i_p,j_p), \nonumber \\
 \end{eqnarray}
 where $\mathfrak{s}_i$ is the representation label at site $i$ and $\mathcal{C}$ is the Hankel contour going counterclockwise around the cut lying between points at $0$ and $\infty$. Here, permutation $\sigma$ relevant for tree-level amplitude $A_{n,k}$, is a permutation of $n$ elements and $k$ is the number of element $i$ with $\sigma (i) < i$. It turns out, that $k$ preserving and $k$ increasing inverse soft operations introduced above within ISL procedure could be conveniently written in terms of $\mathcal{R}$-matrix operators at zero value of spectral parameter $R = \mathcal{R}(0)$:
\begin{eqnarray}
&& \mathcal{S}_{+}(i-1ii+1)A_{n,k}(...,\widehat{(i-1)}^{+},\widehat{(i+1)}^{+},...)
=R_{ii+1}R_{ii-1}A_{n,k}(..., i-1,i+1,...), \nonumber \\
&& \mathcal{S}_{-}(i-1ii+1)A_{n,k}(...,\widehat{(i-1)}^{-},\widehat{(i+1)}^{-},...)
=R_{i+1i}R_{i-1i}A_{n,k}(..., i-1,i+1,...)\delta^{4|4}(\mathcal{W}_i). \nonumber \\
\end{eqnarray}

\subsection{Inverse soft limit and integrability for form factors}
The ISL construction is easy to generalize to the case of form factors. It is expected, that any tree-level form factor in $\mathcal{N}=4$ SYM could be also generated with the use of ISL and the result could be schematically written as \cite{Nandan_GeneretingTreeAmplitudes_ISL}:
\begin{eqnarray}
F(1,2,\ldots ,n|q,\gamma) &=& \sum_{i;L,R}\Big[
\prod_R\mathcal{S}_R F(1,2,\ldots , i, i+1| q,\gamma )|_{subst.} \Big. \nonumber \\
&+& \Big. \prod_L\mathcal{S}_L F(i,i+1,\ldots , n-1, n| q,\gamma )|_{subst.}\Big] ,
\end{eqnarray}
where again soft factors $\mathcal{S}_{L,R}$ are equal to either $\mathcal{S}_+$ or $\mathcal{S}_-$ and $|_{subst.}$ subscript means, that one has to make several folded substitutions as in the amplitude case. In a particular case of form factors of operators from stress tensor multiplet we have
\begin{eqnarray}
Z_{\bold{2},n}(1,...n|q,\gamma)=\sum_{i;R,L}(\prod_{L}\mathcal{S}_L)(\prod_{R}\mathcal{S}_R)Z_{\bold{2},2}|_{subst.}.
\end{eqnarray}
Let us now demonstrate how this construction works on a few particular examples.
MHV sector is trivial.
\begin{figure}[t]
	\begin{center}
		\epsfxsize=11cm
		\epsffile{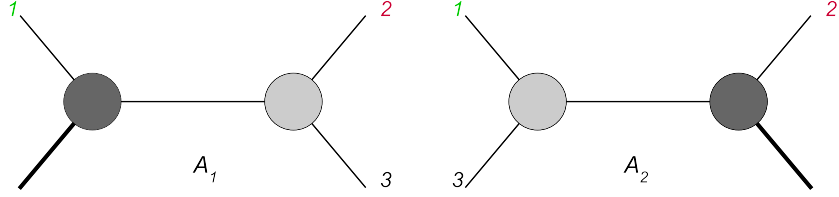}
	\end{center}\vspace{-0.2cm}
	\caption{BCFW diagrams contributing to the $n=3$ case,
		for the $[\textcolor{green}{1},\textcolor{red}{2} \rangle$ shift.
		$A_1=0$ due to  kinematical reasons.}\label{NMHV3}
\end{figure}
\begin{figure}[t]
	\begin{center}
		\epsfxsize=14cm
		\epsffile{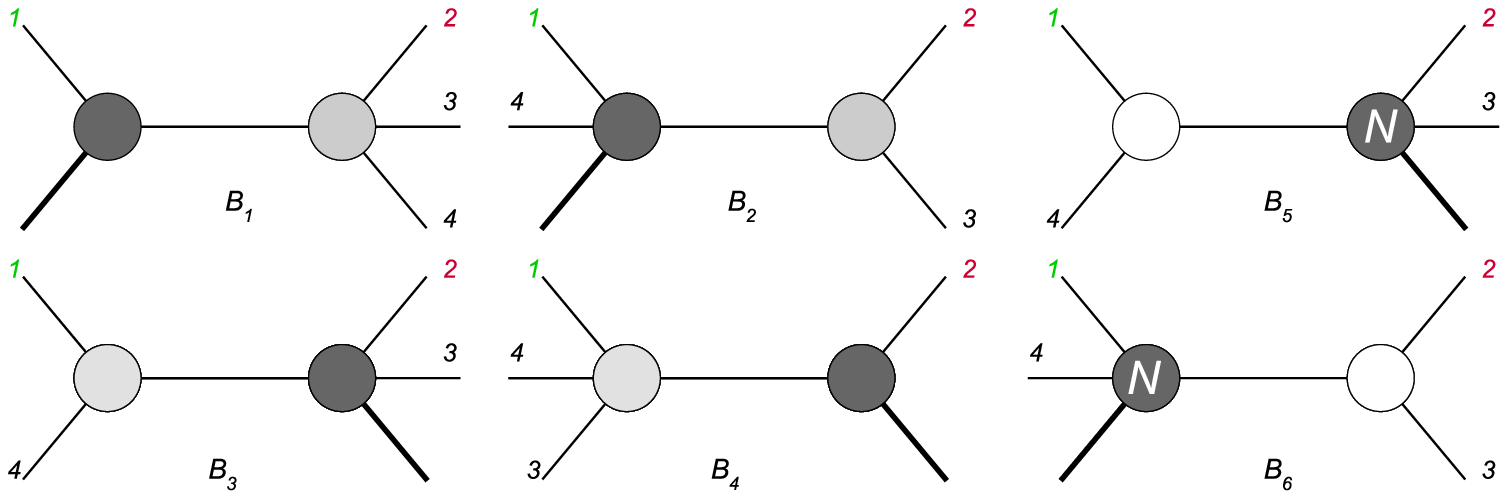}
	\end{center}\vspace{-0.2cm}
	\caption{BCFW diagrams contributing to the $n=4$ case, for the $[\textcolor{green}{1},\textcolor{red}{2} \rangle$ shift.
		$B_{2}=B_{5}=0$ due to kinematical reasons.}\label{NMHV4_12}
\end{figure}
\begin{figure}[t]
	\begin{center}
		\epsfxsize=14cm
		\epsffile{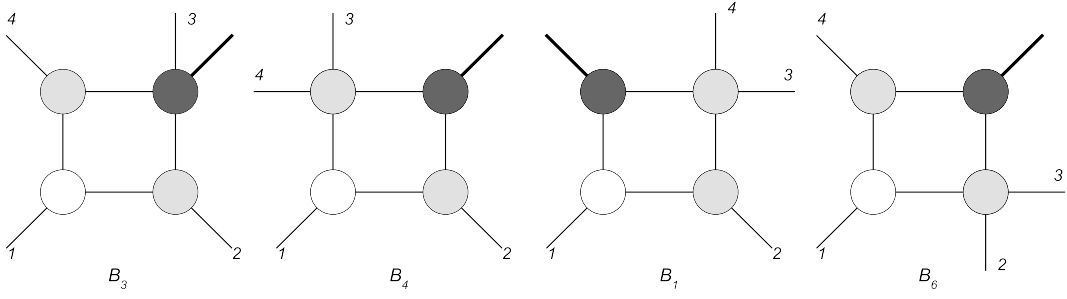}
	\end{center}\vspace{-0.2cm}
	\caption{Schematic representation of the corresponding $R$ functions contributing to the $n=4$ case, for the $[\textcolor{green}{1},\textcolor{red}{2} \rangle$ shift.}\label{NMHV4_12_R}
\end{figure}
So, lets consider NMHV sector with $[1,2 \rangle$ BCFW shift. Recall that $[i-1,i \rangle$ shift is defined as
\begin{eqnarray}
&&\hat{\lambda}_{i}=\lambda_{i}+z\lambda_{i-1},\nonumber\\
&&\hat{\tilde{\lambda}}_{i-1}=\tilde{\lambda}_{i-1}-z\tilde{\lambda}_{i},\nonumber\\
&&\hat{\eta}_{i-1}=\eta_{i-1}+z\eta_{i}.
\end{eqnarray}
In the case of $n=3$ NMHV form factor we have\footnote{Here we use notations and conventions, in particular for $R$-functions from \cite{BORK_POLY}.}:
\begin{eqnarray}
Z_{\bold{2},3}^{NMHV}=A_2=Z_{\bold{2},3}^{MHV}\tilde{R}^{(1)}_{122},
\end{eqnarray}
while for $n=4$ the corresponding expression is given by
\begin{eqnarray}
Z_{\bold{2},4}^{NMHV}=B_1+B_6+B_3+B_4=Z_{\bold{2},4}^{MHV}
\left(R^{(2)}_{142}+\tilde{R}^{(1)}_{133}+R^{(1)}_{132}+\tilde{R}^{(1)}_{122}\right),
\end{eqnarray}
that is
\begin{eqnarray}
~B_1=Z_{\bold{2},4}^{MHV}R^{(2)}_{142},
~B_6=Z_{\bold{2},4}^{MHV}\tilde{R}^{(1)}_{133},
~B_3=Z_{\bold{2},4}^{MHV}R^{(1)}_{132},
~B_4=Z_{\bold{2},4}^{MHV}\tilde{R}^{(1)}_{122}. \nonumber \\
\end{eqnarray}
These expressions containing $R^{(i)}$ functions could be further simplified and we get
\begin{eqnarray}
~A_2(1,2,3)=\delta^8(q_{123}+\gamma)
\frac{\hat{\delta}^4(\eta_1[23]+\eta_2[31]+\eta_3[12])}{q^4[12][23][31]},
\end{eqnarray}
\begin{eqnarray}
~B_1(1,2,3,4)=\delta^8(q_{1234 }+\gamma)
\frac{\hat{\delta}^4(\eta_2[34]+\eta_3[42]+\eta_4[23])}{p^2_{234}[23][34][2|3+4|1\rangle[4|2+3|1\rangle},
\end{eqnarray}
\begin{eqnarray}
B_6(1,2,3,4)=\frac{\delta^8(q_{1234 }+\gamma)}{\langle12\rangle\langle23\rangle\langle34\rangle\langle41\rangle}\frac{1}{q^4}
\frac{\langle23\rangle\langle41\rangle\hat{\delta}^4(X_6)}
{p_{312}^2[4|1+2|3\rangle[4|2+3|1\rangle},\nonumber\\
X_6=\eta_2\langle2|1+3|4]+\eta_3\langle3|1+2|4]+\eta_1\langle1|2+3|4]-\eta_4p_{123}^2,
\end{eqnarray}
\begin{eqnarray}
B_3(1,2,3,4)=\delta^8(q_{1234 }+\gamma)
\frac{\hat{\delta}^4(\eta_4[12]+\eta_1[24]+\eta_2[41])}{p^2_{412}[41][12][4|1+2|3\rangle[2|1+4|3\rangle},
\end{eqnarray}
\begin{eqnarray}
B_4(1,2,3,4)=\frac{\delta^8(q_{1234 }+\gamma)}{\langle12\rangle\langle23\rangle\langle34\rangle\langle41\rangle}\frac{1}{q^4}
\frac{\langle12\rangle\langle23\rangle\hat{\delta}^4(X_6)}
{p_{134}^2[2|1+4|3\rangle[2|3+4|1\rangle},\nonumber\\
X_6=\eta_1\langle1|4+3|2]+\eta_3\langle3|1+4|2]+\eta_4\langle4|1+3|2]-\eta_2p_{134}^2.
\end{eqnarray}
Note, that in fact in all expressions above there is no pole in $q^2$ on the support of $\delta^8(q_{123}+\gamma)$ and $\delta^8(q_{1234}+\gamma)$ functions.

Next, consider the action of $\mathcal{S}_{-}$ inverse soft operation on $Z_{\bold{2},2}^{MHV(0)}$ ($Z_{\bold{2},2}^{MHV(0)}(1,2)\equiv Z_{\bold{2},2}(1,2)$. For example, lets take the following momentum dependence (in $n=3$ all possible combinations of momentum dependence give the same answer because of cyclic symmetry, while in $n=4$ this is no longer the case)
\begin{eqnarray}
\mathcal{S}_{-}(3,1,2)Z_{\bold{2},2}(\hat{3}^{-},\hat{2}^{-})=
\frac{\hat{\delta}^4(\eta_1[23]+\eta_2[31]+\eta_3[12])}{[12][23]^3[31]}
\frac{\delta^8(\hat{q}_2^{-}+\hat{q}_3^{-}+\gamma)}{\langle\hat{2}^{-}\hat{3}^{-}\rangle^2}.
\end{eqnarray}
Here, all momenta and spinor substitutions correspond to $\hat{1}^-$ negative shift. Taking into account, that
\begin{eqnarray}
\delta^8(\hat{q}_2^{-}+\hat{q}_3^{-}+\gamma)\hat{\delta}^4(\eta_1[23]+\eta_2[31]+\eta_3[12])=
\delta^8(q_{123}+\gamma)\hat{\delta}^4(\eta_1[23]+\eta_2[31]+\eta_3[12]),\nonumber\\
\end{eqnarray}
and
\begin{eqnarray}
\langle\hat{2}^{-}\hat{3}^{-}\rangle=([23]\langle23\rangle+[12]\langle12\rangle+[13]\langle13\rangle)/[23]=q^2/[23],
\nonumber\\
\end{eqnarray}
we see that
\begin{eqnarray}
\mathcal{S}_{-}(3,1,2)Z_{\bold{2},2}(\hat{3}^{-},\hat{2}^{-})=A_2(1,2,3)=Z_{\bold{2},3}^{(0)NMHV}.
\end{eqnarray}
In a similar fashion it is easy to obtain, that
\begin{eqnarray}
B_1(1,2,3,4)=\mathcal{S}_{-}(2,3,4)Z_{\bold{2},3}^{MHV}(1,\hat{2}^{-},\hat{4}^{-}) =
\mathcal{S}_{-}(2,3,4)[\mathcal{S}_{+}(\hat{2}^{-},1,\hat{4}^{-})
Z_{\bold{2},2}((\hat{2}^{+})^{-},(\hat{4}^{+})^{-})].\nonumber\\
\end{eqnarray}
Here, the notation $(\hat{2}^{+})^{-}$ means that we have to shift all $\lambda$'s and $\eta$'s with label 2 first according to the substitutions associated with $\mathcal{S}_{+}(2,1,4)$ inverse soft operation ($|^{+}$ superscript) and then shift the result according to the substitutions associated with $\mathcal{S}_{-}(2,3,4)$ inverse soft operation ($|^{-}$ superscript). In general, within ISL procedure one can encounter expressions like $(((((\hat{2}^{+})^{-})^{-})^{-})^{+})^{-}$. Proceeding this way for other $B_i$ functions we get

\begin{eqnarray}
B_3(1,2,3,4)=\mathcal{S}_{-}(2,1,4)Z_{\bold{2},3}^{MHV}(3,\hat{2}^{-},\hat{4}^{-})=
\mathcal{S}_{-}(2,1,4)[\mathcal{S}_{+}(\hat{2}^{-},3,\hat{4}^{-})
Z_{\bold{2},2}((\hat{2}^+)^{-},(\hat{4}^{+})^{-})].\nonumber\\
\end{eqnarray}
\begin{eqnarray}
B_4(1,2,3,4)=\mathcal{S}_{+}(1,4,3)Z_{\bold{2},3}^{NMHV}(3,\hat{1}^{+},\hat{2}^{+})=
\mathcal{S}_{+}(1,4,3)[\mathcal{S}_{-}(\hat{1}^{+},2,\hat{3}^{+})
Z_{\bold{2},2}((\hat{1}^{-})^{+},(\hat{3}^{-})^{+})],\nonumber\\
\end{eqnarray}
\begin{eqnarray}
B_6(1,2,3,4)=\mathcal{S}_{+}(1,2,3)Z_{\bold{2},3}^{NMHV}(4,\hat{1}^{+},\hat{3}^{+})]=
\mathcal{S}_{+}(1,2,3)[\mathcal{S}_{-}(\hat{1}^{+},4,\hat{3}^{+})
Z_{\bold{2},2}((\hat{1}^{-})^{+},(\hat{3}^{-})^{+})].\nonumber\\
\end{eqnarray}
So one sees, that in the case of $n=3,4$ NMHV form factors of stress-tensor operator supermultiplet we can reproduce all BCFW contributions within ISL iterative construction for general $q^2 \neq 0$. It is interesting to note that there are other contributions to BCFW recursion which are equal to zero. Such terms should be annihilated by the corresponding set of $\mathcal{S}_{-}$ and $\mathcal{S}_{+}$ inverse soft operations operators. In addition in $q^2=0$ limit $A_2$ and $B_6,~B_4$  terms are also equal to zero and consequently the combination of inverse soft operations
$\mathcal{S}_{+}(1,2,3)\mathcal{S}_{-}(\hat{1}^{+},4,\hat{3}^{+})$ annihilates $Z_{\bold{2},2}((\hat{1}^{-})^{+},(\hat{3}^{-})^{+})$.

As we already noted in previous section (see also appendices \ref{aB} and \ref{aC}), there are some indications that form factors of the operators, whose anomalous dimensions were studied previously within the context of integrable $PSU(2,2|4)$ spin chain (see for example \cite{BeisertReviewIntegrability}), may exhibit Yangian invariance in the limit $q^2\to 0$. Then, in this case one may wonder what will be the corresponding (analogous to the amplitude case we described before) spin chain description of these form factors. We expect that it will be the same spin chain as in the case of amplitudes with one of the nodes containing a representation space build from $L$ copies of Jordan-Schwinger superoscillators corresponding to twist $L$ operator at that node. A particular simple situation arises in the case of \emph{color-adjoint} form factors. In this case the relative position of operator node in spin chain is fixed (we are considering color-ordered contribution to the form factor) and Lax operator corresponding to the operator node is given by
\begin{eqnarray}
 \mathcal{L} (u,v) = u - v + \sum_{a,b} e_{ab} J_{ba} ~\mbox{where}~
 J_{ab} = \sum_{i=1}^L \overline{\bf w}_{a,i} {\bf w}_{b,i},
 \end{eqnarray}
and $\overline{\bf w}_{a,i}, {\bf w}_{b,i}$ are $L$ copies of superoscillators used to describe twist $L$ operator. All the other Bethe ansatz machinery should be similar to the case of scattering amplitudes. We suppose to return to this question in one of our future publications.

\section{Conclusion}\label{p6}
In this article, we derived soft theorems for the form factors from 1/2-BPS and Konishi supermultiplets in $\mathcal{N}=4$ SYM at tree level and considered one loop corrections to such theorems on several particular examples. In $\mathcal{N}=4$ SYM at tree and loop level the soft theorems have the same form both in the case of form factors and amplitudes. Soft theorems are independent from the specific choice of operator or from the presence of UV divergences related to operator. In the case, when momentum carried by operator becomes soft the behavior of form factors are regular.
We have also presented a possible Grassmannian integral representation of form factors, which was checked on a few simple examples. It was shown, that in the $q^2\to 0$ limit ($q$ is the form factor momentum) there is some evidence in favor of Yangian invariance of tree-level form factors of single trace operators studied previously in the context of $PSU(2,2|4)$ $\mathcal{N}=4$ SYM spin chain. We have checked on a few simple examples, that the inverse soft limit (ISL) iterative procedure works also in the case of form factors and commented on the applicability of quantum inverse scattering method (QISM) for the  construction of Yangian invariants relevant for form factors in the limit of $q^2\to 0$.

\section*{Acknowledgements}
The authors would like to thank D. I. Kazakov for valuable and stimulating discussions.
Financial support of RFBR grant \# 14-02-00494 and contract \# 02.А03.21.0003 from 27.08.2013 with Russian Ministry of Science and Education is kindly acknowledged.

\appendix

\section{$\mathcal{N}=4$ harmonic superspaces conventions}\label{aA}

The $\mathcal{N}=4$ harmonic superspace is obtained by adding additional bosonic
coordinates (harmonic variables) to the $\mathcal{N}=4$ coordinate
superspace or on-shell momentum superspace. These additional bosonic coordinates parameterize the coset
\begin{equation}
\frac{SU(4)}{SU(2) \times SU(2)' \times U(1)}
\end{equation}
and carry the  $SU(4)$ index $A$, two copies of the $SU(2)$ indices $a,a'$ and the $U(1)$ charge $\pm$
\begin{equation}
(u^{+a}_{A},~u^{-a'}_A)~\mbox{and their conjugate ones}~(\bar{u}^{A}_{+a},~\bar{u}^{A}_{-a'}).
\end{equation}
Using these variables all the Grassmann objects with
$SU(4)_R$ indices could be rewritten in terms of Grassmann objects with $SU(2) \times SU(2)' \times U(1)$ indices. This way. each index of the (anti)fundamental $SU(4)$ representation $A$ splits into $A = (+a,-a')$, where $\pm$ indicates the $U(1)$ charge and $a,a'=1,2$ are $SU(2)\times SU(2)'$ indices. The Grassmann coordinates in the original $\mathcal{N}=4$ coordinate superspace are transformed as
\begin{eqnarray}
&& \theta^{+a}_{\alpha} = u^{+a}_{A}\theta^A_{\alpha},~~
\theta^{-a'}_{\alpha} = u^{-a'}_{A}\theta^A_{\alpha},\\
&& \bar{\theta}_{+a\dot{\alpha}} = \bar{u}^{A}_{+a}\bar{\theta}_{A\dot{\alpha}},~~
\bar{\theta}_{-a'\dot{\alpha}}=\bar{u}^{A}_{-a'}\bar{\theta}_{A\dot{\alpha}},
\end{eqnarray}
and in the opposite direction\footnote{Here we use notations indentical to \cite{HarmonyofFF_Brandhuber}, which are slightly different from
\cite{BORK_POLY}. One can convert one notation into another using
mnemonic rule $X_{+a, here}=X^-_{a, there}$, $X_{-a', here}=X^+_{a', there}$,
$X^{+a, here}=X^{+a, there}$,$X^{-a', here}=X^{-a', there}$.}
\begin{equation}
\theta^A_{\alpha}=\theta^{+a}_{\alpha}\bar{u}^{A}_{+a}+
\theta^{-a'}_{\alpha}\bar{u}^{A}_{-a'},
\end{equation}
\begin{equation}
\bar{\theta}_{A\dot{\alpha}}=\bar{\theta}_{+a\dot{\alpha}}u^{+a}_A+
\bar{\theta}_{-a'\dot{\alpha}}u^{-a'}_{A}.
\end{equation}
The same is true for supercharges:
\begin{equation}
Q_{A\alpha}\rightarrow(Q_{+a\alpha},~Q_{-a'\alpha}),
~\bar{Q}^{A}_{\dot{\alpha}}\rightarrow(\bar{Q}^{+a}_{\dot{\alpha}},~\bar{Q}^{-a'}_{\dot{\alpha}}),
\end{equation}
where
\begin{eqnarray}
Q_{A\alpha} = \frac{\partial}{\partial\theta^{A\alpha}} -
\bar{\theta}^{\dot{\alpha}}_{A}q_{\alpha\dot{\alpha}}, \quad
\bar{Q}^{A}_{\dot{\alpha}} = -
\frac{\partial}{\partial\bar{\theta}^{\dot{\alpha}}_A} + \theta^{A\alpha}q_{\alpha\dot{\alpha}} .
\end{eqnarray}

So the $\mathcal{N}=4$ harmonic superspace is parameterized with the
following set of coordinates
\begin{eqnarray}
\mbox{$\mathcal{N}=4$ harmonic
superspace}&=&\{x^{\alpha\dot{\alpha}},
~\theta^{+a}_{\alpha},
~\theta^{-a'}_{\alpha},
~\bar{\theta}_{+a\dot{\alpha}},
~\bar{\theta}_{-a'\dot{\alpha}}
~u
\}.\nonumber\\
\end{eqnarray}
Harmonic variables can also be introduced in on-shell momentum superspace used to treat on-shell states of the theory on equal footing with operators from operator supermultiplets. Using harmonic variables one can write:
\begin{eqnarray}
\mbox{$\mathcal{N}=4$ harmonic
on-shell momentum superspace}&=&\{\lambda_{\alpha},\tilde{\lambda}_{\dot{\alpha}},~\eta_{+a},\eta_{-a'},~u\}.
\nonumber\\
\end{eqnarray}
Here $\lambda_{\alpha}$ and $\tilde{\lambda}_{\dot{\alpha}}$ are the $SL(2,\mathbb{C})$ spinors associated with momentum carried by a massless state (particle): $p_{\alpha\dot{\alpha}}=\lambda_{\alpha}\tilde{\lambda}_{\dot{\alpha}}$, $p^2=0$. Supercharges acting on this n-particle on-shell momentum superspace can be written as
\begin{eqnarray}
q_{+a\alpha}=\sum_{i=1}^n\lambda_{\alpha,i}\eta_{+a,i},
~~q_{-a'\alpha}=\sum_{i=1}^n\lambda_{\alpha,i}\eta_{-a',i},
\end{eqnarray}
and
\begin{eqnarray}
\bar{q}^{+a}_{\dot{\alpha}}=\sum_{i=1}^n\tilde{\lambda}_{\dot{\alpha},i}\frac{\partial}{\partial\eta_{+a,i}},
~~\bar{q}^{-a'}_{\dot{\alpha}}=\sum_{i=1}^n\tilde{\lambda}_{\dot{\alpha},i}\frac{\partial}{\partial\eta_{-a',i}}.
\end{eqnarray}

The Grassmann delta functions, which one can encounter in this paper, are given by
($ \langle ij \rangle\equiv\lambda_{\alpha}^i\lambda^{j\alpha}$):
\begin{eqnarray}
\delta^{-4}(q_{+a\alpha})=\sum_{i,j=1}^n\prod_{a,b=1}^2\langle ij \rangle \eta_{+a,i}\eta_{+b,j},~~
\delta^{+4}(q_{-a'\alpha})=\sum_{i,j=1}^n\prod_{a',b'=1}^2\langle ij \rangle
\eta_{-a',i}\eta_{-b',j},
\end{eqnarray}
\begin{eqnarray}
\hat{\delta}^{-2}(X_{+a})=\prod_{a=1}^2X_{+a},~~\hat{\delta}^{+2}(X_{-a'})=\prod_{a=1}^2X_{-a'}.
\end{eqnarray}
We will also use the following abbreviations
\begin{eqnarray}
\delta^{-4}\delta^{+4}\equiv\delta^{8},~\hat{\delta}^{-2}\hat{\delta}^{+2}\equiv\hat{\delta}^4.
\end{eqnarray}
With the help of these delta functions one can rewrite the $\mbox{MHV}_3$ and $\overline{\mbox{MHV}}_3$ amplitudes, $R_{rst}$
functions etc. in the form nearly identical to
the form they have in the ordinary on-shell momentum superspace.

The Grassmann integration measures for the on-shell momentum superspace are defined as
\begin{eqnarray}
d^{-2}\eta=\prod_{a=1}^2d\eta_{+a},
~d^{+2}\eta=\prod_{a=1}^2d\eta_{-a'},
~d^{2}\eta d^{-2}\eta\equiv d^4\eta.
\end{eqnarray}
and for the case of ordinary superspace as
\begin{eqnarray}
d^{-4}\theta=\prod_{a,\alpha=1}^2d\theta_{+a\alpha},
~d^{+4}\theta=\prod_{a',\alpha=1}^2d\theta_{-a'\alpha},
\end{eqnarray}
Note also, that $\delta^{\pm4}$ functions can be conveniently  represented as product of two $\hat{\delta}^{\pm2}$
functions using the identity (here we drop the $SU(2)$ and $SL(2,\mathbb{C})$ indices),
\begin{equation}\label{supersumm_delta_to_hatt_deltas}
\delta^{\mp4}(q_{\pm})=\langle lm
\rangle^{2}\hat{\delta}^{\mp2}\left(\eta_{\pm,l}+\sum_{i=1}^n\frac{\langle
mi \rangle}{\langle ml
\rangle}\eta_{\pm,i}\right)\hat{\delta}^{\mp2}\left(\eta_{\pm,m}+\sum_{i=1}^n\frac{\langle
li \rangle}{\langle lm \rangle}\eta_{\pm,i}\right), ~i \neq l,~i\neq m.
\end{equation}
and subsequently integrated as usual Grassmann delta functions.

\section{MHV form factors of operators from stress tensor supermultiplet with $q^2=0$ and dual conformal invariance}\label{aB}
As was mentioned earlier in the main text, the representation of amplitudes/form factors in terms of on-shell diagrams is tightly related to their symmetry properties. Scattering amplitudes (at least at tree level) are Yangian invariant for general kinematics.
Yangian algebra appears in this context as a closure of two algebras of ordinary and dual (super)conformal symmetry transformations. Here, we are going to speculate about properties of form factors with respect to the dual (super)conformal transformations.

\begin{figure}[t]
    \begin{center}
        \leavevmode
        \epsfxsize=12cm
        \epsffile{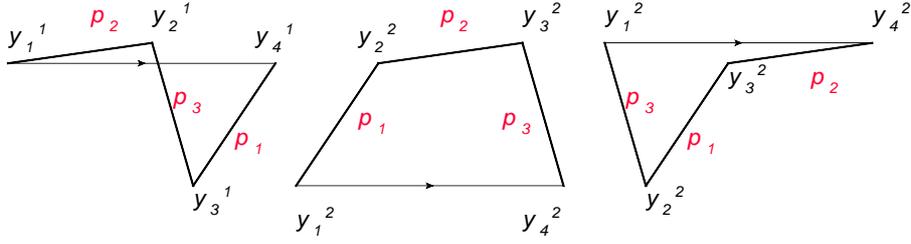}
    \end{center}\vspace{-0.2cm}
    \caption{Different closed contours for $n=3$.}\label{DifferentContours}
\end{figure}

\begin{figure}[t]
    \begin{center}
        \leavevmode
        \epsfxsize=12cm
        \epsffile{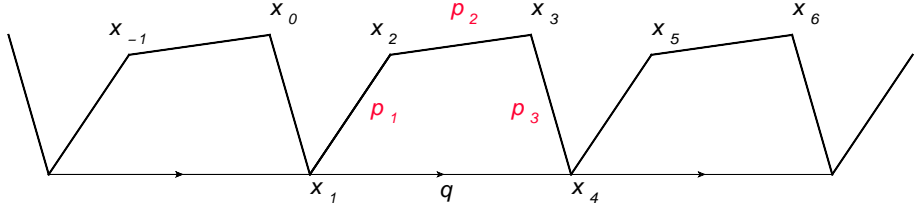}
    \end{center}\vspace{-0.2cm}
    \caption{Periodical contour for $n=3$.}\label{PeriodicalContour}
\end{figure}

Let us focus on MHV tree-level form factors of operators
from self dual part of stress tensor supermultiplet,
$q^2=0$, $q=\lambda_q\tilde{\lambda}_q$ case.
\begin{equation}
    Z_{\bold{2},n}^{MHV}(1,...,n)=
    \delta^{4}(\sum_{i=1}^{n}\lambda_i\tilde{\lambda}_i-\lambda_q\tilde{\lambda}_q)
    \frac{\delta^{-4}(\sum_{i=1}^n\lambda_i\eta_{+,i} + \gamma_+)
        \delta^{+4}(\sum_{i=1}^n\lambda_i\eta_{-,i}+ \gamma_-)}
    {\langle 12\rangle\langle23\rangle...\langle
        n1\rangle},\nonumber\\
\end{equation}
where we have to put $\gamma_-=0$ in the end. Set of momenta $\{\lambda_i\tilde{\lambda}_i,\lambda_q\tilde{\lambda}_q\}$ forms closed contour, all elements of which lie on the lightcone.
To describe with dual variables $y_i^{\alpha\dot{\alpha}}$, defined as
\begin{equation}
    p_i^{\alpha\dot{\alpha}}=y_i^{\alpha\dot{\alpha}}-y_{i+1}^{\alpha\dot{\alpha}},
\end{equation}
all kinematical invariants one can encounter
in form factor computation it is necessary to consider different
closed contours \cite{Brandhuber_FormFactorsN4SYM,HarmonyofFF_Brandhuber,BORK_POLY,Zhiboedov_Strong_coupling_FF}
$\Gamma_k$ where momentum $q$ is inserted at different positions among
$p_i$ momenta (see Fig. \ref{DifferentContours}). It is convenient to parametrize these contours with different sets of coordinates $y^{k}_i$:
\begin{equation}
    \delta^{4}(\sum_{i=1}^{n}\lambda_i\tilde{\lambda}_i-\lambda_q\tilde{\lambda}_q)=
    \frac{1}{2n}\sum_k\sum_{i=1}^{n}
    \delta^{4}(y_{i}^k-y_{i+n}^k).
\end{equation}
These $y_{i}^k$ have well defined conformal weights and their
momentum conservation delta functions will transform covariantly with respect to  dual conformal inversions. They also obey linear relations
similar to ($n=3$ case, see Fig. \ref{DifferentContours})
\begin{equation}
    y_3^{1}-y_4^{1}=y_1^{2}-y_2^{2}=p_1,
\end{equation}
So, one can think of $y_{i}^k$ as of points on large periodical contour $x_i$
\cite{Brandhuber_FormFactorsN4SYM,HarmonyofFF_Brandhuber,BORK_POLY,Zhiboedov_Strong_coupling_FF}.
Sum over $k$ runs through all inequivalent contours $\Gamma_k$.
In principle, one can think about $y^k_i$
and $x_i$ coordinates as equivalent on the periodical contour (see Figs. \ref{DifferentContours} and \ref{PeriodicalContour}).

The same discussion as above is also valid for fermionic counterparts of
$y_i^k$ dual coordinates $\theta_i^k$, defined as
\begin{eqnarray}
    \lambda_{\alpha,i}\eta_{+a,i}&=&\theta_{+a\alpha,i}-\theta_{+a\alpha,i+1},
    \nonumber\\
    \lambda_{\alpha,i}\eta_{-a',i}&=&\theta_{-a'\alpha,i}-\theta_{-a'\alpha,i+1}.
\end{eqnarray}

Another question, related to the dual conformal properties of MHV form factor, is how to define the generators of dual conformal transformations
(for example, supercharges $Q^{\pm, dual}$) acting on the operator  variables
$\{\lambda_q,\tilde{\lambda}_q,\gamma^+,\gamma^-\}$. One can assume, that after the introduction of dual variables, the action of dual (super)conformal generators will be identical to their action of on the i-th particle. Indeed, the operator
is now parameterized by the same number of variables, at least in the bosonic sector,\footnote{In the fermionic sector we have to take $\gamma_-\to 0$ limit, but one should always keep in mind, that this limit is an artifact of dealing with the chiral truncation of stress tensor operator supermultiplet and for the full operator supermultiplet $\gamma_-\ne 0$.}
$$
\{\lambda_{\alpha,q};~\tilde{\lambda}_{\dot{\alpha},q};~
\gamma_{+a,\alpha}=\lambda_{\alpha,q}\eta_{+a,q};~
\gamma_{-a',\alpha}=\lambda_{\alpha,q}\eta_{-a',q}\},
$$
as the external particle
$$
\{\lambda_{\alpha,i};~\tilde{\lambda}_{\dot{\alpha},i};~\lambda_{\alpha,i}\eta_{+a,i};~\lambda_{\alpha,i}\eta_{-a',i}\},
$$
and the naive interpretation of dual (super)conformal transformations in the simplest case of $Q^{dual,+}$ generator as ``something that
acts on $\gamma_+$ as if it is the ordinary superspace coordinate''
\cite{DualConfInvForAmplitudesCorch}, that is
\begin{equation}
    Q^{dual,+}=\frac{\partial}{\partial\gamma_+},
\end{equation}
supports this conjecture. In addition to these  two questions there is another subtlety related to the fact, that we are dealing with chiral truncation of stress tensor operator supermultiplet and consequently not all supercharges will annihilate tree-level form factor $Z_{\bf{2},n}^{MHV}$. This could be avoided by considering full non-chiral stress tensor operator supermultiplet. Here, however, for simplicity we will restrict ourselves to the $SU(2)$ $R$-symmetry invariant subsector, where all supercharges annihilate $Z_{\bf{2},n}^{MHV}$ form factor. So, in what follows, we are using the above prescriptions regarding both the structure of dual coordinates (we are using $y_i^k$) and the form of generators acting on variables which parametrize operator (they are the same as for the $i$-th external particle). As was already discussed in section \ref{p4}, we can rewrite MHV form factor of stress tensor operator supermultiplet\footnote{Here for saving space we will use abbreviation $\{\lambda_i,\tilde{\lambda}_i,\eta_i\}\equiv i$.} (let's take $n=2$ for example) as
\begin{eqnarray}
    Z_{\bold{2},2}^{MHV}(1,2)=[S(1,q,2)]^{-1}A_{3,0}(1,2,q).
\end{eqnarray}
Here, $A_{3,0}(1,2,q)$ is Yangian invariant and is annihilated\footnote{Except at collinear configurations of external momenta.}
by all generators $J^{\mathbf{AB}}$ of (super)conformal and dual (super)conformal
$J^{\mathbf{AB},(1)}$ algebras ($\mathbf{A}$ is multi-index for $\alpha$, $\dot{\alpha}$ and
$+ a$, $-a'$) atleast before taking $\gamma_-=0$ limit. Ordinary (super)conformal transformations of form factors where already considered in \cite{Wilhelm_AmplitudesFormFactorsDilatationOperator}, so here we
restrict ourself to the case of dual (super)conformal transformations $J^{\mathbf{AB},(1)}$. $Z_{\bold{2},2}^{MHV}(1,2)$
is annihilated by $J^{\mathbf{AB},(1)}$ if
\begin{eqnarray}
    [J^{\mathbf{AB},(1)},[S(1,q,2)]^{-1}]=0.
\end{eqnarray}
The only problem may come from generators, that contain terms like $\partial/\partial\lambda_i$,
and there is only one such generator $K^{\alpha\dot{\alpha}}$ - generator of dual special conformal transformations. More accurately, if we define $K^{\alpha\dot{\alpha}}$ as
\cite{DualConfInvForAmplitudesCorch}
\begin{eqnarray}
    K^{\alpha\dot{\alpha}}=\sum_i\left(
x_i^{\dot{\beta}\alpha}x_i^{\dot{\alpha}\beta}\frac{\partial}{\partial x_i^{\dot{\beta}\beta}}
+x_i^{\dot{\alpha}\beta}\theta^{\Lambda,\alpha}_i\frac{\partial}{\partial \theta^{\Lambda,\beta}_i}
+x^{\dot{\alpha}\beta}_i\lambda^{\alpha}_i\frac{\partial}{\partial \lambda^{\beta}_i}
+x^{\dot{\beta}\alpha}_{i+1}\tilde{\lambda}_i^{\dot{\alpha}}\frac{\partial}{\partial \tilde{\lambda}^{\dot{\beta}}_i}
+\tilde{\lambda}^{\dot{\beta}}_i\theta^{\Lambda,\alpha}_{i+1}\frac{\partial}{\partial{\eta^{\Lambda}_i}}\right),\nonumber\\
\end{eqnarray}
then, the action of this generator on $A_{3,0}(1,2,q)$ is given by ($x_3\equiv x_q$)
\begin{eqnarray}
K^{\alpha\dot{\alpha}}A_{3,0}(1,2,q)=-(x_1^{\alpha\dot{\alpha}}+x_2^{\alpha\dot{\alpha}}
+x_q^{\alpha\dot{\alpha}})A_{3,0}(1,2,q).
\end{eqnarray}
In the amplitude case one redefines $K^{\alpha\dot{\alpha}}$  generator up to the terms proportional to unity operator to absorb $\sum_{i=1}^{n}x_i^{\alpha\dot{\alpha}}$ contribution and get
\begin{eqnarray}
K^{\alpha\dot{\alpha}}A_{n,0}(1,2,q)=0.
\end{eqnarray}
Here, we will not use this redefinition, but proceed considering the action of this generator on form factor instead. Note also, that $K^{\alpha\dot{\alpha}}$ generator defined here
is written in the "universal form". If the amplitude or form factor are written in terms of the on shell momentum superspace
variables, then only 3 last terms are relevant. If on the other hand amplitude or form factor are written in terms of dual variables, then only first 2 terms contribute. Returning to the action of the $K^{\alpha\dot{\alpha}}$  generator on form factors we get
\begin{eqnarray} K^{\alpha\dot{\alpha}}Z_{\bold{2},2}^{MHV}(1,2)=-(\sum_{i=1}^{2}x_i^{\alpha\dot{\alpha}}+x_q^{\alpha\dot{\alpha}})Z_{\bold{2},2}^{MHV}(1,2)+
    (K^{\alpha\dot{\alpha}}[S(1,q,2)]^{-1})S(1,q,2)Z_{\bold{2},2}^{MHV}(1,2),\nonumber\\
\end{eqnarray}
One can easily see, that the only terms in $K^{\alpha\dot{\alpha}}$ contributing to $(K^{\alpha\dot{\alpha}}[S(1,q,2)]^{-1})S(1,q,2)$ are the terms proportional to
$\partial/\partial \lambda^{\beta}_i$. Using the relation
\begin{eqnarray}
x^{\alpha\dot{\alpha}}=
\frac{x^{\beta\dot{\alpha}}\lambda_{i,\beta}\lambda_{j}^{\alpha}}{\langle ji\rangle}
+\frac{x^{\beta\dot{\alpha}}\lambda_{j,\beta}\lambda_{i}^{\alpha}}{\langle ij\rangle},~\mbox{for linear independent}~\lambda_{i}^{\alpha},\lambda_{j}^{\alpha},
\end{eqnarray}
it is easy to see, that
\begin{eqnarray}
(K^{\alpha\dot{\alpha}}[S(1,q,2)]^{-1})S(1,q,2)=x_q^{\alpha\dot{\alpha}},
\end{eqnarray}
and
\begin{eqnarray} K^{\alpha\dot{\alpha}}Z_{\bold{2},2}^{MHV}(1,2)=-(\sum_{i=1}^{2}x_i^{\alpha\dot{\alpha}})Z_{\bold{2},2}^{MHV}(1,2).
\end{eqnarray}
The same result could be obtained directly from the action of
$K^{\alpha\dot{\alpha}}$ on $Z_{\bold{2},2}^{MHV}(1,2)$.
The generalization to the case of $Z_{\bold{2},n}^{MHV}(1,\ldots,n)$ is trivial:
\begin{eqnarray}
  K^{\alpha\dot{\alpha}}Z_{\bold{2},n}^{MHV}(1,\ldots,n)=
  -(\sum_{i=1}^{n}x_i^{\alpha\dot{\alpha}})Z_{\bold{2},n}^{MHV}(1,\ldots,n).
\end{eqnarray}

So we see, that if the action of generators of dual (super)conformal
transformations  on the operator in form factor with $q^2=0$
have the same form as their action on
external particles (which is very likely) then MHV form factors transforms covariantly with respect to dual (super)conformal transformations. This fact and the form of $\mbox{N}^k\mbox{MHV}$
tree level form factors (see appendix \ref{aC}) strongly suggest the presence of dual conformal symmetry (and possibly Yangian symmetry) at
least in the case of tree-level form factors of stress tensor operator supermultiplet with $q^2=0$.

\section{BCFW for form factors of stress tensor supermultiplet with $q^2=0$, $\mbox{N}^k\mbox{MHV}$ sector}\label{aC}
Let us consider the BCFW recursion for NMHV form factor of stress tensor operator supermultiplet in momentum supertwistor notation \cite{BORK_POLY}. Performing the shift of
momentum supertwistor as \cite{Henrietta_Amplitudes,Arkani_Hamed_Integrahd}
\begin{eqnarray}
\hat{\mathcal{Z}}_2=\mathcal{Z}_2+w\mathcal{Z}_3,
\end{eqnarray}
which is equivalent to the $[1,2\rangle$ shift in the momentum superspace and considering the contour integral
\begin{eqnarray}
\oint \frac{d w}{w}\hat{Z}_{\bold{2},n}^{N^kMHV}(w)=0,
\end{eqnarray}
we get the following
recursion relations in the case of NMHV sector:
\begin{eqnarray}\label{q20_NMHV_BCFW_Twistors}
    &&\frac{Z_{\bold{2},n}^{NMHV}}{Z_{\bold{2},n}^{MHV}}(\mathcal{Z}_{2-n},...,\mathcal{Z}_{1},\mathcal{Z}_{2},\mathcal{Z}_{3},...,\mathcal{Z}_{1+n})=
    \frac{Z_{\bold{2},n-1}^{NMHV}}{Z_{\bold{2},n-1}^{MHV}}(\mathcal{Z}_{2-n},...,\mathcal{Z}_{1},\mathcal{Z}_{3},\mathcal{Z}_{4},...,\mathcal{Z}_{1+n})+
    \nonumber\\
    &+&\sum_{j=3}^n[1,2,3,j,j+1]+\sum_{j=3}^{n-1}[1,2,3,j-n,j+1-n].
\end{eqnarray}
The momentum supertwistor is defined as
\begin{eqnarray}
    \mathcal{Z}_i^{\mathbf{A}}=\left(
    \begin{array}{ccc}
        Z_{i}^M \\
        \chi_{+a/-a',i}
    \end{array}
    \right),
\end{eqnarray}
where the fermionic part of the supertwistor $\chi$ can be written as:
\begin{eqnarray}
    \chi_{+a,i}=\theta_{+a,i}\lambda_{i},
    ~\chi_{-a',i}=\theta_{-a',i}\lambda_{i},
\end{eqnarray}
while the bosonic part is given by
\begin{eqnarray}
    Z_i^{M}=\left(
    \begin{array}{ccc}
        \lambda_{i}^{\alpha} \\
        \mu^{\dot{\alpha}}_i
    \end{array}
    \right),~
    \mu^{\dot{\alpha}}_i=x^{\alpha\dot{\alpha}}_i\lambda_{\alpha i}.
\end{eqnarray}
Here, index $M$ stands for Lorentz indexes $\alpha$ and $\dot{\alpha}$ and $x^{\alpha\dot{\alpha}}_i,\theta_{+a,i},\theta_{-a,i}$ are dual variables defined on the periodical contour. $[a,b,c,d,e]$ is the
rational function of 5 twistor variables $Z_a,...,Z_e$
and their supersymmetric counterparts
\begin{eqnarray}
    [a,b,c,d,e]=\frac{\hat{\delta}^4(\langle a,b,c,d\rangle\chi_e+\mbox{cycl.})}
    {\langle a,b,c,d\rangle\langle b,c,d,e\rangle\langle c,d,e,a\rangle\langle d,e,a,b\rangle
        \langle e,a,b,c\rangle}.
\end{eqnarray}
So, for example, from Eq. (\ref{q20_NMHV_BCFW_Twistors}) (similar to \cite{BORK_POLY})  we get
\begin{eqnarray}
\frac{Z_{\bold{2},4}^{NMHV}}{Z_{\bold{2},4}^{MHV}}=[1,2,3,4,5]+[1,2,3,0,-1],
\end{eqnarray}
\begin{eqnarray}
\frac{Z_{\bold{2},5}^{NMHV}}{Z_{\bold{2},5}^{MHV}}
&=&[1,3,4,5,6]+[1,2,3,4,5]+[1,2,3,5,6]\nonumber\\
&+&[-1,0,1,3,4]+[1,2,3,-2,-1]+[1,2,3,-1,0].
\end{eqnarray}
The obtained expressions are Yangian invariant for general twistor configuration and we have used the standard notation for dual conformal $SU(2,2)$ invariant
\begin{eqnarray}
\langle i,j,k,l\rangle=\varepsilon_{M_1M_2M_3M_4}Z^{M_1}_iZ^{M_2}_jZ^{M_3}_kZ^{M_4}_l.
\end{eqnarray}
where $\varepsilon_{M_1M_2M_3M_4}$ is totally antisymmetric tensor.

Note, that to incorporate all kinematical invariants for the $n$-point
form factor one needs $2n$ twistor variables living on the periodical contour. So, here we use the set of
$(\mathcal{Z}_{2-n},...,\mathcal{Z}_{1+n})$ twistors to describe
$n$-point form factor \cite{HarmonyofFF_Brandhuber,BORK_POLY} at least in NMHV sector. It should be noted, that one can rewrite
$R^{(1)}_{1st}$, $R^{(2)}_{1st}$ functions using momentum twistors as:
\begin{eqnarray}
R^{(1)}_{1st}=[1,t,t+1,s-n,s+1-n],
~
R^{(2)}_{1st}=[1,t,t+1,s,s+1].
\end{eqnarray}
In general $\mbox{N}^k\mbox{MHV}$ sector the similar BCFW recursion relations read:
\begin{eqnarray}
    &&\frac{Z_{\bold{2},n}^{N^kMHV}}{Z_{\bold{2},n}^{MHV}}
    (...,\mathcal{Z}_{-n+2},\mathcal{Z}_{-n+3},...,\mathcal{Z}_{1},\mathcal{Z}_{2},\mathcal{Z}_{3},
    ...,\mathcal{Z}_{n},\mathcal{Z}_{n+1},...)=
    \nonumber\\
    &=&\frac{Z_{\bold{2},n-1}^{N^kMHV}}{Z_{\bold{2},n-1}^{MHV}}(...,\mathcal{Z}_{1-n},...,\mathcal{Z}_{1},\mathcal{Z}_{3},\mathcal{Z}_{4},...,\mathcal{Z}_{1+n},...)
    \nonumber\\
    &+&\sum_{j=3}^n[1,2,3,j,j+1]\times\frac{A_{n_1,k_1}}{A_{n_1,0}}
    \left(\mathcal{Z}_I,\hat{\mathcal{Z}}_{2},...,\mathcal{Z}_{j}\right)\times
    \frac{Z_{\bold{2},n_2}^{N^{k_2}MHV}}{Z_{\bold{2},n_2}^{MHV}}
    \left(...,\mathcal{Z}_{0},\mathcal{Z}_{1},\mathcal{Z}_I,\mathcal{Z}_{j+1},...\right)
    \nonumber\\
    &+&\sum_{j=3}^{n-1}[1,2,3,j-n,j+1-n]\times\frac{Z_{\bold{2},n_1}^{N^{k_1}MHV}}{Z_{\bold{2},n_1}^{MHV}}
    \left(...,\mathcal{Z}_{j-n},\mathcal{Z}_I,\hat{\mathcal{Z}}_{2},\mathcal{Z}_{3},...\right)
    \times\frac{A_{n_2,k_2}}{A_{n_2,0}}
    \left(\mathcal{Z}_I,\mathcal{Z}_{1},...,\mathcal{Z}_{j+1-n}\right),
    \nonumber\\
\end{eqnarray}
where\footnote{$(jj+1)\bigcap(klm)=\mathcal{Z}_j\langle j+1klm\rangle
    +\mathcal{Z}_{j+1}\langle jklm\rangle$}
\begin{eqnarray}
    \mathcal{Z}_I=(jj+1)\bigcap(123)~\mbox{and}~\hat{\mathcal{Z}}_2=(12)\bigcap(0jj+1),
\end{eqnarray}
\begin{eqnarray}
    n_1+n_2-2=n,~k_1+k_2+1=k.
\end{eqnarray}
Thus, in $q^2\to 0$ limit the recursion relations are given by the same formula as in $q^2\neq0$ case,
but without terms with coefficients $c^{m}_n$ (see \cite{BORK_POLY} for details).
It is assumed that one has to put $\chi_{-,i\pm n}=\chi_{-,i}$ (which is momentum twistor equivalent
of $\gamma_-=0$ condition) in all expressions above.

\end{document}